# Van der Waals engineering of ferromagnetic semiconductor heterostructures for spin and valleytronics

Ding Zhong[1†], Kyle L. Seyler[1†], Xiayu Linpeng[1†], Ran Cheng[2], Nikhil Sivadas[2], Bevin Huang[1], Emma Schmidgall[1], Takashi Taniguchi[3], Kenji Watanabe[3], Michael A. McGuire[4], Wang Yao[5], Di Xiao[2], Kai-Mei C. Fu[1,6], Xiaodong Xu[1,7*]

[1]Department of Physics, University of Washington, Seattle, Washington 98195, USA.

[2]Department of Physics, Carnegie Mellon University, Pittsburg, Pennsylvania 15213, USA.

[3]National Institute for Materials Science, 1-1 Namiki, Tsukuba 305-0044, Japan.

[4]Materials Science and Technology Division, Oak Ridge National Laboratory, Oak Ridge, Tennessee, 37831, USA.

[5]Department of Physics and Center of Theoretical and Computational Physics, University of Hong Kong, Hong Kong, China.

[6]Department of Electrical Engineering, University of Washington, Seattle, Washington 98195, USA.

[7]Department of Materials Science and Engineering, University of Washington, Seattle, Washington 98195, USA.

[†]These authors contributed equally to this work.
[*]Correspondence to: xuxd@uw.edu

**Abstract**

The integration of magnetic material with semiconductors has been fertile ground for fundamental science as well as of great practical interest toward the seamless integration of information processing and storage. Here we create van der Waals heterostructures formed by an ultrathin ferromagnetic semiconductor $CrI_3$ and a monolayer of $WSe_2$. We observe unprecedented control of the spin and valley pseudospin in $WSe_2$, where we detect a large magnetic exchange field of nearly 13 T and rapid switching of the $WSe_2$ valley splitting and polarization via flipping of the $CrI_3$ magnetization. The $WSe_2$ photoluminescence intensity strongly depends on the relative alignment between photo-excited spins in $WSe_2$ and the $CrI_3$ magnetization, due to ultrafast spin-dependent charge hopping across the heterostructure interface. The photoluminescence detection of valley pseudospin provides a simple and sensitive method to probe the intriguing domain dynamics in the ultrathin magnet, as well as the rich spin interactions within the heterostructure.

## Main Text

Manipulating exchange interactions in magnetic heterostructures has proven to be an effective way to engineer highly functional materials. For example, control of the interlayer exchange coupling (*1–3*) and exchange bias (*4*) has enabled widely used magnetic storage technologies such as giant magnetoresistance (*2*, *3*) and magnetic tunnel junctions (*5*). Currently there is intense focus to expand the class of functional materials that can utilize magnetism via proximity exchange effects. Heterostructures of magnetic material integrated with superconductors (*6*), topological insulators (*7*, *8*), and semiconductors (*9*) have been proposed and investigated. The incorporation of magnetism with semiconductor electronics is particularly enticing for spintronic applications (*10*, *11*). For example, exchange interactions between the ferromagnet spins and semiconductor charge carriers lead to a tunable spin polarization (*12–14*). Our work extends this field into new territory by constructing and exploring the exchange interactions in a van der Waals interface between a monolayer $WSe_2$ semiconductor and an ultrathin ferromagnetic semiconductor $CrI_3$.

The use of two-dimensional (2D) materials to form magnetic heterostructures has the advantage that the proximity-induced exchange interaction, usually considered as a perturbative effect for bulk materials, can fundamentally alter the electronic structure of the 2D material. Theoretical work has predicted strong exchange effects (*15*, *16*) and emergent phenomena (*17–21*) in systems that integrate 2D materials with bulk, 3D magnets. Proximity-induced exchange effects have been experimentally demonstrated in graphene/EuS heterostructures (*22*, *23*) and recently, valley-polarized electroluminescence from 2D semiconductors has been achieved (*24*, *25*). In this latter case, the degree of polarization remains modest and the effect of exchange fields has not been observed. Furthermore, due to the polycrystalline nature and small grain size of the 3D magnets in these devices, external magnetic fields are required to polarize the ferromagnets for any observable magnetic functionality.

Compared to the approach of interfacing 2D materials with 3D magnets, a van der Waals heterostructure has several advantages (*26–28*). Lattice mismatch is not an issue, thus minimizing chemical modification and interfacial damage, which is desirable for engineering a clean interface for optimal interactions. Since single crystals are used, the twist angle and stacking order between constituent 2D materials becomes a potentially powerful control knob, enabling for instance, the ability to engineer and study magnetic multilayer van der Waals stacks with unique spin textures. The flexibility of the layer stacking process also facilitates the creation of van der Waals heterostructures between layered ferromagnets and a diverse set of other 2D materials, such as graphene, Weyl semimetals ($WTe_2$), superconductors ($NbSe_2$), and beyond.

Here we report that a van der Waals ferromagnetic heterostructure formed by monolayer $WSe_2$ and ultrathin $CrI_3$ enables unprecedented control of spin and valley pseudospin in monolayer $WSe_2$, while revealing a new platform for studying the dynamics of layered magnets. Figure 1A displays the heterostructure architecture, where vertically stacked monolayer $WSe_2$ and ~10 nm $CrI_3$ (Fig. 1B) are sandwiched by thin (10-20 nm) hexagonal boron nitride (h-BN) flakes to prevent sample degradation (see Materials and Methods). We present results from two devices (12 and 9 nm $CrI_3$), which show similar behavior, and a top view of device 2 is shown in Fig 1C. In our design, we use monolayer semiconducting $WSe_2$ due to its well-established spin and valley pseudospin properties (Fig. 1D, dashed lines), including circularly polarized valley optical selection rules (*29*), spin-valley locking effects (*30*), and valley Zeeman splitting of about 0.22 meV/T (*31–34*). $CrI_3$ (Fig. 1B) is a layered ferromagnetic semiconductor with an optical gap near 1.2 eV (*35*). Bulk $CrI_3$ crystal exhibits a Curie temperature ($T_C$) of 61 K, a saturation magnetization of 3.1 $\mu_B$ per Cr, and an out-of-plane easy axis (*36*). The last feature is important because in $WSe_2$ the spin quantization axis is also out-of-plane, with opposite orientations in opposite valleys due to the spin-valley locking effect (Fig. 1D). Thus, only perpendicular effective magnetic fields can lift the valley degeneracy.

The sample is cooled in a helium cryostat and excited by a HeNe laser (1.96 eV). Below, we label the polarized photoluminescence (PL) spectra as $P_1P_2$, where $P_1$ ($P_2$) represents the excitation (detection) polarization, which can be either right (R, σ+) or left (L, σ-) circularly polarized. Figure 1E (left panel) displays a typical PL spectrum of device 1 at a temperature of 65 K, above the $T_C$ of $CrI_3$. We attribute the PL to a positively charged trion state in $WSe_2$ due to the type-II band alignment, with the $CrI_3$ conduction band lying below that of $WSe_2$ (supplementary S1, S2). Here the RR and LL spectra are nearly identical, as expected from time-reversal symmetry between the valleys in $WSe_2$.

Remarkably, the valley degeneracy is lifted as the sample is cooled below the $T_C$ of $CrI_3$. Figure 1E (right panel) shows a representative spectrum taken at 5 K. The RR spectrum exhibits both a larger peak intensity and energy than LL. The extracted valley splitting between RR and LL spectra is about 3.5 meV (see supplementary section S3 for fitting details), equivalent to an effective magnetic field of about 13 T (supplementary section S4). Further measurements reveal that the magnitude of the valley splitting is independent of the excitation power (supplementary section S5), which rules out carrier density effects as a dominant mechanism for the valley splitting. Thus, we attribute the energy splitting to a strong magnetic exchange field between the $CrI_3$ spins and $WSe_2$ excitons. We emphasize that since the optical transition is determined by the energy difference between the conduction and valence bands, the observed valley splitting reflects the distinct coupling of the exchange field with the conduction and valence bands (Fig. 1D, solid lines). This is due to the different orbital character of the conduction and valence band edges (*19*, *20*).

We further study the polarized PL while sweeping an applied magnetic field **B** perpendicular to the sample plane (Faraday geometry, **B** parallel to easy-axis, Fig. 1A). In Fig 2A, the total PL intensity (I) for L excitation ($I_{LL}+I_{LR}$) and R excitation ($I_{RR}+I_{RL}$) is plotted as a function of applied magnetic field and emission energy for device 1. Six distinct regimes are denoted by the arrows, with three sharp jumps in PL intensity and peak energy near 0 T and ±0.85 T, and two smaller jumps near ±1.85 T. Figure 2B illustrates the overlaid RR and LL spectra at the six selected magnetic fields corresponding to each regime, highlighting the multiple jumps in peak energy and intensity. In Figs. 2C and 2D, we display the valley splitting and the normalized difference between RR and LL intensities, $\rho = \frac{I_{RR}-I_{LL}}{I_{RR}+I_{LL}}$, as a function of increasing (orange curve) and decreasing (green green) applied magnetic field. Hysteresis loops, a hallmark of ferromagnetic effects, are observed at $B = \pm 0.85$ T and $B = 0$ T. The widths of observed hysteresis loops are around 50 mT. Similar measurements in Voigt geometry (**B** parallel to the sample plane) demonstrate a rotation of the magnetization from in-plane to out-of-plane as the applied magnetic field is reduced to zero, confirming the out-of-plane easy axis in ultrathin $CrI_3$ (supplementary section S6). In the following discussions, we ignore the direct influence of the applied magnetic field on $WSe_2$ PL intensity and energy (*31*, *32*, *37*, *38*), since the exchange field is much stronger than the applied field (typically < ~1 T).

A notable observation in Fig. 2A is that at a fixed applied magnetic field, the total PL intensity depends strongly on the excitation helicity. To understand this phenomenon, we focus on the regime with $B > 0.9$ T, where the $CrI_3$ magnetization **M** is nearly aligned with **B**. As shown in Fig.

1D, R and L excitation create $|K,\uparrow>$ and $|-K,\downarrow>$ electrons, respectively, where ↑ (↓) represents the electron spin orientation. In $CrI_3$, the lowest energy unoccupied conduction bands are mainly composed of the Cr atom's spin-polarized $e_g$ orbitals, with spin orientation parallel to the ground state spin (supplementary section S1). Therefore, electron hopping from $WSe_2$ to $CrI_3$ is only allowed for the aligned spin $|K,\uparrow>$ (Fig. 2E). Under the condition that optical absorption at 1.96 eV is independent of light helicity (supplementary section S7), the electron spins created by R excitation have an extra non-radiative relaxation channel compared to L excitation. This results in strongly quenched PL as well as a broader spectral linewidth compared to L excitation. The helicity-dependent PL is similar to that observed recently in an epitaxially grown ferromagnet/quantum well heterostructure (*39*), however the PL polarization and intensity modulation is an order of magnitude stronger in our system with an order of magnitude smaller applied magnetic field range (supplementary section S8).

In Fig. 2F, we plot the extracted linewidth and PL intensity of the LL spectrum versus applied magnetic field. The general trend is that the linewidth broadens whenever the PL intensity decreases, and vice versa. This is further evidence supporting the proposed physical picture that in the spin-aligned configuration, the allowed charge hopping opens up an additional non-radiative relaxation channel. We can estimate the hopping rate by the difference in spectral width between RR and LL conditions. For $B$ = 1 T, the difference is ~5 meV, implying a spin-dependent charge hopping rate on the order of $\frac{\hbar}{5\ \mathrm{meV}} \approx 130\ fs$ , where  is the reduced Planck constant. Moreover, the spectral widths vary with the applied magnetic field (supplementary section S9), suggesting that the hopping rate can be tuned by controlling the $CrI_3$ magnetization.

Another outstanding feature in Figs. 2C and 2D is the sharp change of the valley splitting and ρ near $B$ = ±0.85 T and 0 T (movie S1). For example, when sweeping the applied magnetic field down through the transition near $B$ = -0.85 T, the valley splitting changes from -1.3 meV to 3.4 meV in a span of about 30 mT (Fig. 2C). This corresponds to a valley splitting rate of over 150 meV/T, nearly three orders of magnitude larger than can be achieved by the Zeeman effect in bare $WSe_2$. Moreover, when sweeping through -0.8 T in the other direction and also near zero field, the switching is even faster, occurring within a ≤ 6 mT span (supplementary section S10). We also note that the critical fields are independent of the sweep rate (supplementary section S11). The sign change of the valley splitting and ρ implies a flip of the $CrI_3$ magnetization. This behavior is expected for a ferromagnet when the applied magnetic field crosses zero. However, the sign change near ±0.85 T is intriguing since the field has not changed direction. This complicated magnetic field dependence implies rich magnetization dynamics in the ultrathin $CrI_3$ substrate. By comparison, bulk $CrI_3$ exhibits the expected magnetization loop at zero magnetic field with small remanence and coercivity (*36*).

Since both the valley splitting and ρ are tightly connected to the $CrI_3$ magnetization, spatial maps of these parameters should reveal the underlying magnetic domain structure. Inspired by this idea, we performed spatially resolved, polarized PL measurements as a function of applied magnetic field on device 2 (boxed region in Fig. 1C). Figure 3 displays spatial maps of ρ as a function of magnetic field sweeping direction, which are ordered to highlight time-reversal pairs (supplementary section S12 for valley splitting maps). The time-reversal pair consists of two maps acquired at opposite applied magnetic field while sweeping in opposite direction. We find that each time reversal pair shows excellent consistency in their pattern except for the opposite sign, validating the stability of the system. The domain structure and ρ disappear above ~60 K (supplementary section S13), indicating that $T_C$ is similar to bulk $CrI_3$ (*36*).

These maps reveal the evolution of the domain structure between -1.1 to 1.1 T in the ultrathin $CrI_3$. When the applied magnetic field is larger than the coercive field, the uniform color across the entire heterostructure indicates full magnetization of $CrI_3$. When the applied magnetic field is set at an intermediate value, a multi-domain structure appears with a domain size on the order of a few microns. We define domains that only switch signs with the reversal of the applied field as strong domains (indicated by the red arrow in $B$ = -0.5 T map), while those that flip without switching the applied field as weak domains (indicated by the blue arrow). Unlike the strong domains, the weak domains flip three times within one sweep. Therefore, if the laser spot is on a weak domain, the corresponding valley splitting and ρ will also flip three times within one sweep, which is consistent with the observation in Figs. 2C and 2D. A plausible interpretation of the multiple flips is the competition between the Zeeman energy and the magnetic dipole energy between strong and weak domains. In a certain applied magnetic field range, the dipolar interaction exceeds the Zeeman energy in a weak domain, which causes it to flip to reduce the magnetic dipole energy. A simple model of the magnetic domain evolution, which considers the Zeeman energy, the dipolar interaction between strong and weak domains, and the easy-axis anisotropy, qualitatively reproduces the observed curves in Fig. 2C (supplementary materials S14).

Measurement of the valley-polarized exciton PL in $WSe_2$ additionally provides a new local probe to investigate spin and domain dynamics in adjacent ferromagnets through van der Waals engineering. Figure 4 shows multiple distinct ρ-$B$ curves at different sample positions. As expected from the spatial maps in Fig. 3, if the laser is focused on a weak domain (Fig. 4B and C), the magnetization can flip without changing the applied magnetic field direction, while in a strong domain (Fig. 4E), there is only one hysteresis loop centered at zero field. In addition to coarse changes in magnetization, fine structure is also observed. For instance, in Fig. 4C we observe fine steps of ρ as the applied magnetic field varies. At first glance, the observation is an echo of the Barkhausen effect due to the rapid change of the domain size in ferromagnets. However, the laser spot is towards the center of the weak domain, which is less likely to be affected by changes in domain size. Furthermore, the ρ-$B$ curves near the domain boundary do not exhibit these fine steps, as seen in Fig. 4D. One possibility to consider is whether the magnetization in thin $CrI_3$ can flip layer by layer, as in certain types of layered antiferromagnets (*40*, *41*). Detailed $CrI_3$ thickness dependence will be crucial to help understand the electronic and magnetic coupling beyond the heterostructure interface. In addition, Fig. 4D displays an overshoot of ρ (or spikes) which may be related to negative differential magnetization (*42*). Further interesting topics include unraveling the origin of the slow jumps at $\pm$1.85T, which manifest themselves as an opposite trend between Fig. 2C and 2D, and may indicate a different origin from the other jumps at lower fields.

These intriguing phenomena will require additional study to elucidate the underlying mechanisms. It is already evident, however, that this probe is highly sensitive to spin interactions and magnetization dynamics, providing a powerful addition to conventional techniques such as magneto-optical Kerr rotation in studying magnetism in thin films. In addition, the integration of van der Waals magnetic materials into other heterostructures should empower researchers to harness exchange interactions and interfacial effects for exploring novel physical phenomena and spintronics at the atomically thin limit.

## Materials and Methods

### Device fabrication

Bulk crystals of $WSe_2$, $CrI_3$, and h-BN were first exfoliated onto 90 nm $SiO_2$ on Si. Due to the instability of $CrI_3$ in air, we performed exfoliation in a glovebox with $O_2$ and $H_2O$ levels below 0.5 ppm. Immediately after finding the $CrI_3$ sample, we assembled the heterostructure stack (h-BN/$CrI_3$/monolayer $WSe_2$/h-BN) using a polycarbonate-based transfer technique (43) in a glovebox. The h-BN thickness is about 20 nm and $CrI_3$ is ~10 nm. The chloroform rinse was performed in ambient environment for 2 minutes. We note that bare $CrI_3$ flakes readily hydrate in air, decomposing within seconds. There were no signs of degradation in the h-BN sandwiched sample in ambient conditions for at least an hour, allowing us to transport and mount the sample for measurement.

### Photoluminescence

The samples were measured in a continuous helium flow cryostat with a 7 T superconducting magnet, which can operate in both Faraday and Voigt geometry. Photoluminescence (PL) measurements were performed in reflection geometry using continuous-wave excitation from a HeNe laser (1.96 eV) that was power-stabilized (30 μW) and focused to ~1 μm with an aspheric lens. The back-reflected PL was collected by the same lens and detected using a spectrometer and Si CCD. Samples were measured at 5 K and in Faraday geometry unless otherwise specified. For the magnetic-field-dependent measurements, we continuously scanned the magnetic field and collected a full hysteresis sweep (up and down in applied magnetic field) for each polarization configuration before switching to the next polarization. Good overlap at saturation magnetic fields indicated the stability of the measurement and allowed comparison of magnetic-field-dependent data between different polarizations.

A two-axis piezoelectric scanning mirror was employed to scan the laser spot over the sample for spatially resolved measurements. Liquid crystal variable waveplates were used to allow repeatable and quick (< 100 ms) switching between different circular polarizations for excitation and collection. Thus, we took both RR and LL polarization configurations at each pixel (1 second integration for each polarization) before moving the laser, which rules out sample drift effects in our data. As shown in Fig. 3 in the main text, there is evident time-reversal symmetry between R and L polarized excitation versus applied magnetic fields. For magnetic-field-dependent data at selected sample positions in Fig. 4, we simply took RR data and time-reversed it to get expected LL data, then compared them to determine the expected ρ.

## References

1. M. D. Stiles, Interlayer exchange coupling. *J. Magn. Magn. Mater.* **200**, 322–337 (1999).
2. M. N. Baibich *et al.*, Giant magnetoresistance of (001)Fe/(001)Cr magnetic superlattices. *Phys. Rev. Lett.* **61**, 2472–2475 (1988).
3. G. Binasch, P. Grünberg, F. Saurenbach, W. Zinn, Enhanced magnetoresistance in layered magnetic structures with antiferromagnetic interlayer exchange. *Phys. Rev. B.* **39**, 4828–4830 (1989).
4. A. E. Berkowitz, K. Takano, Exchange anisotropy — a review. *J. Magn. Magn. Mater.* **200**, 552–570 (1999).
5. J.-G. Zhu, C. Park, Magnetic tunnel junctions. *Mater. Today.* **9**, 36–45 (2006).
6. A. I. Buzdin, Proximity effects in superconductor-ferromagnet heterostructures. *Rev. Mod. Phys.* **77**, 935–976 (2005).
7. C. Lee, F. Katmis, P. Jarillo-Herrero, J. S. Moodera, N. Gedik, Direct measurement of proximity-induced magnetism at the interface between a topological insulator and a ferromagnet. *Nat. Commun.* **7**, 12014

(2016).
8. F. Katmis *et al.*, A high-temperature ferromagnetic topological insulating phase by proximity coupling. *Nature*. **533**, 513–516 (2016).
9. A. H. MacDonald, P. Schiffer, N. Samarth, Ferromagnetic semiconductors: moving beyond (Ga,Mn)As. *Nat Mater*. **4**, 195–202 (2005).
10. J. Fabian, C. Matos-Abiague, P. Ertler, I. Stano, I. Zutic, Semiconductor spintronics. *Acta Phys. Slov.* **57**, 565 (2007).
11. G. A. Prinz, Hybrid ferromagnetic-semiconductor structure. *Science* **250**, 1092-1097 (1990).
12. R. I. Dzhioev, B. P. Zakharchenya, P. A. Ivanov, V. L. Korenov, Detection of the magnetization of a ferromagnetic film in a Ni/GaAs structure from the polarization of electrons of the semiconductor. *JETP Lett.* **60**, 650–654 (1994).
13. V. L. Korenev *et al.*, Long-range p-d exchange interaction in a ferromagnet-semiconductor hybrid structure. *Nat. Phys*. **12**, 85–91 (2016).
14. R. C. Myers, A. C. Gossard, D. D. Awschalom, Tunable spin polarization in III-V quantum wells with a ferromagnetic barrier. *Phys. Rev. B*. **69**, 161305 (2004).
15. H. X. Yang *et al.*, Proximity effects induced in graphene by magnetic insulators: first-principles calculations on spin filtering and exchange-splitting gaps. *Phys. Rev. Lett.* **110**, 46603 (2013).
16. H. Haugen, D. Huertas-Hernando, A. Brataas, Spin transport in proximity-induced ferromagnetic graphene. *Phys. Rev. B*. **77**, 115406 (2008).
17. Z. Qiao *et al.*, Quantum anomalous Hall effect in graphene from Rashba and exchange effects. *Phys. Rev. B*. **82**, 161414 (2010).
18. Z. Qiao *et al.*, Quantum anomalous Hall effect in graphene proximity coupled to an antiferromagnetic insulator. *Phys. Rev. Lett.* **112**, 116404 (2014).
19. J. Qi, X. Li, Q. Niu, J. Feng, Giant and tunable valley degeneracy splitting in $MoTe_2$. *Phys. Rev. B*. **92**, 121403 (2015).
20. Q. Zhang, S. A. Yang, W. Mi, Y. Cheng, U. Schwingenschlögl, Large spin-valley polarization in monolayer $MoTe_2$ on top of EuO(111). *Adv. Mater.* **28**, 959–966 (2016).
21. W. Han, R. K. Kawakami, M. Gmitra, J. Fabian, Graphene spintronics. *Nat. Nano*. **9**, 794–807 (2014).
22. P. Wei *et al.*, Strong interfacial exchange field in the graphene/EuS heterostructure. *Nat. Mater*. **15**, 711–716 (2016).
23. Z. Wang, C. Tang, R. Sachs, Y. Barlas, J. Shi, Proximity-induced ferromagnetism in graphene revealed by the anomalous Hall effect. *Phys. Rev. Lett.* **114**, 16603 (2015).
24. O. L. Sanchez, D. Ovchinnikov, S. Misra, A. Allain, A. Kis, Valley polarization by spin injection in a light-emitting van der Waals heterojunction. *Nano Lett.* **16**, 5792–5797 (2016).
25. Y. Ye *et al.*, Electrical generation and control of the valley carriers in a monolayer transition metal dichalcogenide. *Nat. Nano*. **11**, 598–602 (2016).
26. A. K. Geim, I. V Grigorieva, Van der Waals heterostructures. *Nature*. **499**, 419–425 (2013).
27. K. S. Novoselov, A. Mishchenko, A. Carvalho, A. H. C. Neto, O. Road, 2D materials and van der Waals heterostructures, *Science* **353** aac9439 (2016).
28. J. Zhang, B. Zhao, Y. Yao, Z. Yang, Robust quantum anomalous Hall effect in graphene-based van der Waals heterostructures. *Phys. Rev. B*. **92**, 165418 (2015).
29. D. Xiao, G.-B. Liu, W. Feng, X. Xu, W. Yao, Coupled spin and valley physics in monolayers of $MoS_2$ and other group-VI dichalcogenides. *Phys. Rev. Lett.* **108**, 196802 (2012).
30. X. Xu, W. Yao, D. Xiao, T. F. Heinz, Spin and pseudospins in layered transition metal dichalcogenides. *Nat Phys*. **10**, 343–350 (2014).
31. G. Aivazian *et al.*, Magnetic control of valley pseudospin in monolayer $WSe_2$. *Nat. Phys*. **11**, 148–152 (2015).
32. A. Srivastava *et al.*, Valley Zeeman effect in elementary optical excitations of monolayer $WSe_2$. *Nat Phys*. **11**, 141–147 (2015).
33. D. MacNeill *et al.*, Breaking of valley degeneracy by magnetic field in monolayer $MoSe_2$. *Phys. Rev. Lett.* **114**, 37401 (2015).
34. Y. Li *et al.*, Valley splitting and polarization by the Zeeman effect in monolayer $MoSe_2$. *Phys. Rev. Lett.* **113**, 266804 (2014).
35. J. F. Dillon, H. Kamimura, J. P. Remeika, Magneto-optical properties of ferromagnetic chromium trihalides. *J. Phys. Chem. Solids*. **27**, 1531–1549 (1966).
36. M. A. McGuire, H. Dixit, V. R. Cooper, B. C. Sales, Coupling of crystal structure and magnetism in the

layered, ferromagnetic insulator $CrI_3$. *Chem. Mater.* **27**, 612–620 (2015).
37. G. W. and L. B. and M. M. G. and T. A. and E. L. I. and E. P. and X. M. and B. Urbaszek, Magneto-optics in transition metal diselenide monolayers. *2D Mater.* **2**, 34002 (2015).
38. A. A. Mitioglu *et al.*, Optical investigation of monolayer and bulk tungsten diselenide ($WSe_2$) in high magnetic fields. *Nano Lett.* **15**, 4387–4392 (2015).
39. V. L. Korenev *et al.*, Dynamic spin polarization by orientation-dependent separation in a ferromagnet–semiconductor hybrid. *Nat. Commun.* **3**, 959 (2012).
40. O. Hellwig, T. L. Kirk, J. B. Kortright, A. Berger, E. E. Fullerton, A new phase diagram for layered antiferromagnetic films. *Nat. Mater.* **2**, 112–116 (2003).
41. O. Hellwig, A. Berger, E. E. Fullerton, Domain walls in antiferromagnetically coupled multilayer films. *Phys. Rev. Lett.* **91**, 197203 (2003).
42. Y. Yao, H. C. Mireles, J. Liu, Q. Niu, J. L. Erskine, Negative differential magnetization in ultrathin Fe on vicinal W(100). *Phys. Rev. B.* **67**, 174409 (2003).
43. P. J. Zomer, M. H. D. Guimarães, J. C. Brant, N. Tombros, B. J. van Wees, Fast pick up technique for high quality heterostructures of bilayer graphene and hexagonal boron nitride. *Appl. Phys. Lett.* **105** (2014).
44. J. P. Perdew, K. Burke, M. Ernzerhof, Generalized gradient approximation made simple. *Phys. Rev. Lett.* **77**, 3865–3868 (1996).
45. J. P. Perdew, K. Burke, M. Ernzerhof, Generalized gradient approximation made simple. *Phys. Rev. Lett.* **78**, 1396 (1997).
46. P. Giannozzi *et al.*, QUANTUM ESPRESSO: a modular and open-source software project for quantum simulations of materials. *J. Phys. Condens. Matter.* **21**, 395502 (2009).
47. S. Grimme, Semiempirical GGA-type density functional constructed with a long-range dispersion correction. *J. Comput. Chem.* **27**, 1787–1799 (2006).
48. V. Barone *et al.*, Role and effective treatment of dispersive forces in materials: Polyethylene and graphite crystals as test cases. *J. Comput. Chem.* **30**, 934–939 (2009).
49. A. M. Jones *et al.*, Optical generation of excitonic valley coherence in monolayer $WSe_2$. *Nat. Nano.* **8**, 634–638 (2013).
50. S. Esho, Anomalous magneto-optical hysteresis loops of sputtered Gd-Co films. *Jpn. J. Appl. Phys.* **15**, 93 (1976).

## Acknowledgments

**Funding:** This work is mainly supported by the Department of Energy, Basic Energy Sciences, Materials Sciences and Engineering Division (DE-SC0008145 and SC0012509), and University of Washington Innovation Award. WY is supported by the Croucher Foundation (Croucher Innovation Award), the RGC of Hong Kong (HKU17305914P), and the HKU ORA. Work at ORNL (MAM) was supported by the US Department of Energy, Office of Science, Basic Energy Sciences, Materials Sciences and Engineering Division. KW and TT acknowledge support from the Elemental Strategy Initiative conducted by the MEXT, Japan and a Grant-in-Aid for Scientific Research on Innovative Areas "Science of Atomic Layers" from JSPS. XX, DX, and KF acknowledges the support a Cottrell Scholar Award. XX acknowledges the support from the State of Washington funded Clean Energy Institute and from the Boeing Distinguished Professorship in Physics.

**Author contributions:** XX conceived the project. DZ and BH fabricated the sample. KS, DZ, and XL performed the experiment, assisted by ES and supervised by XX and KCF. MAM synthesized and characterized the bulk $CrI_3$ crystal. TT and KW synthesized and characterized the bulk boron nitride crystal. RC, NS, DX and WY provided the theoretical support. XX, KS, DZ, and DX wrote the paper with input from all authors. All authors discussed the results.

**Competing interests:** The authors declare no competing interests.

## Figures

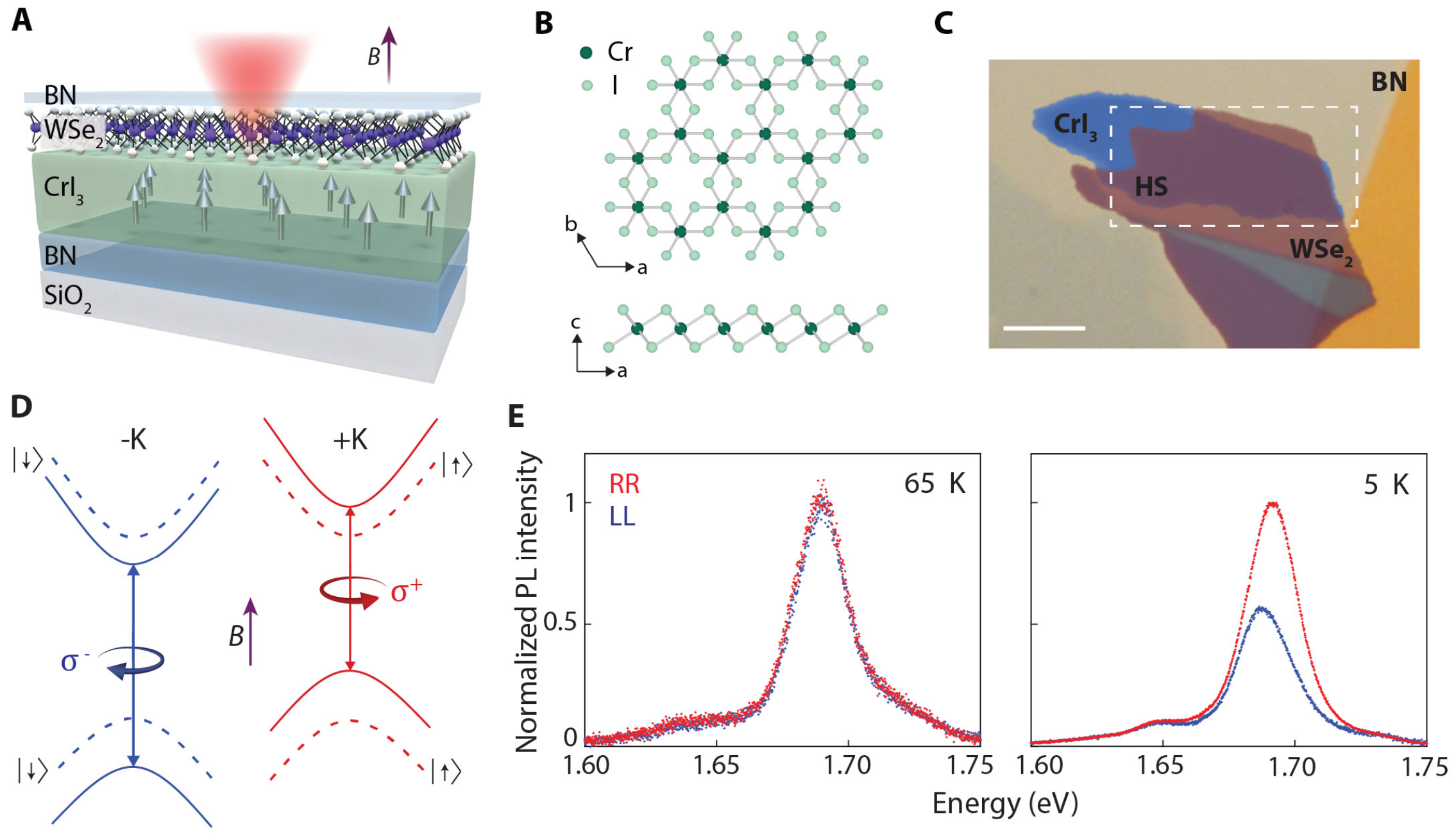

**Fig. 1**. **Ultrathin $CrI_3$/$WSe_2$ heterostructure and observation of spontaneous magnetization.** (**A**) Schematic of van der Waals heterostructure formed by monolayer $WSe_2$ and ferromagnetic layered semiconductor $CrI_3$, and encapsulated by h-BN. (**B**) Top and side view of $CrI_3$ crystal structure. (**C**) Optical microscope image of device 2. The $WSe_2$/$CrI_3$ heterostructure is sandwiched by optically transparent h-BN. Scale bar, 5 μm. (**D**) Spin-valley locking effect and valley-dependent optical selection rules in monolayer $WSe_2$. Dashed (solid) lines indicate the band edges before (after) exchange field coupling. Black arrows denote spins. (**E**) Circularly polarized photoluminescence spectra above $T_C$ (65 K, left) and below $T_C$ (5 K, right) in the absence of an applied magnetic field. It is evident that the valley degeneracy is lifted at 5 K due to the magnetic proximity effect.

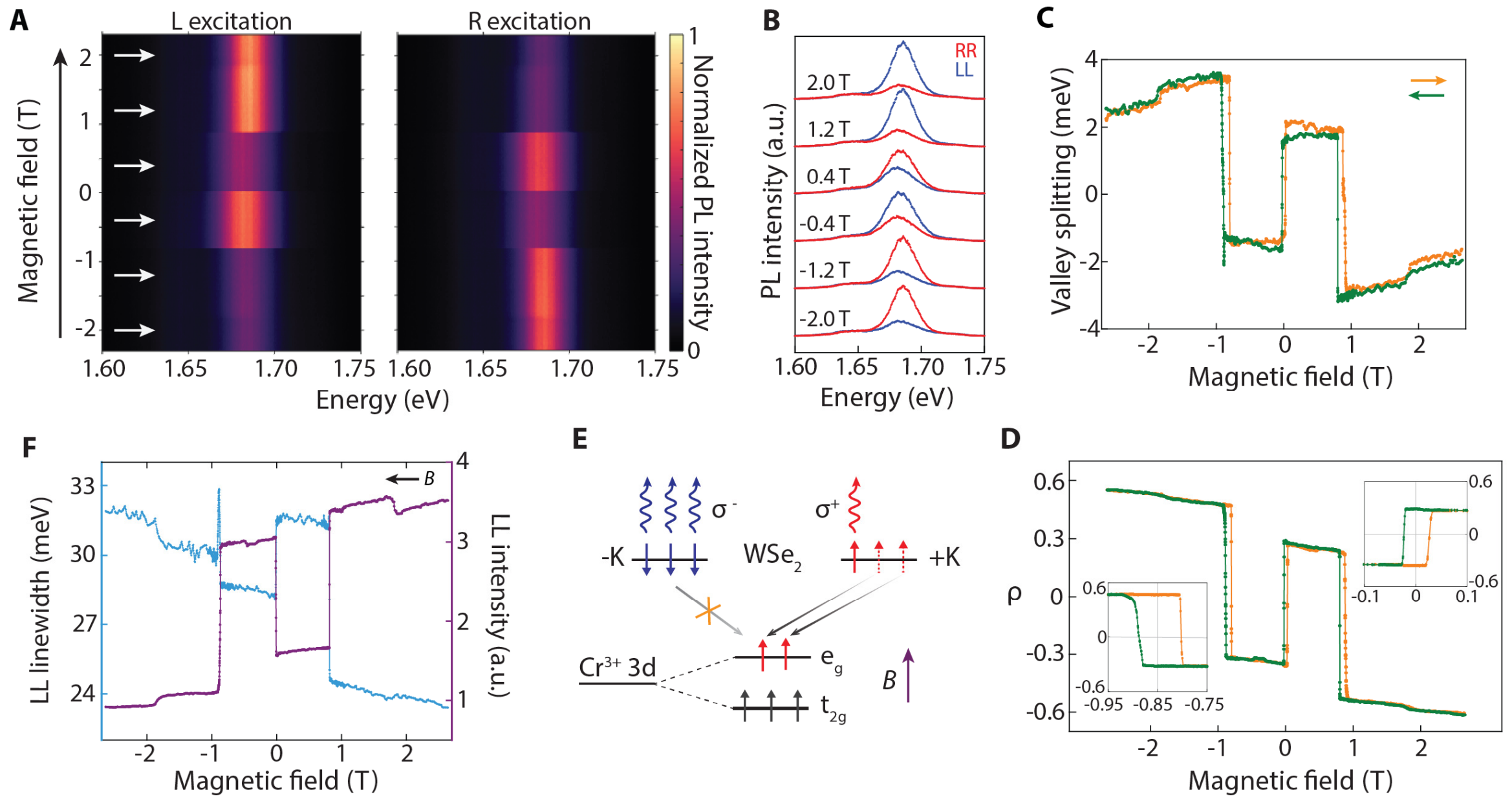

**Fig. 2**. **Ferromagnetic substrate control of spin and valley pseudospin dynamics.** (**A**) Maps of the total photoluminescence intensity as a function of emission energy and applied magnetic field for Left circular (L) and Right circular (R) excitation. The black arrow indicates the applied magnetic field sweeping direction. (**B**) RR and LL spectra at selected magnetic fields (indicated by the white arrows in Fig. 2A). See text for definition of RR and LL. (**C**) Valley splitting and (**D**) normalized PL intensity difference between RR and LL (ρ) as a function of applied magnetic field sweeping up (orange) and down (green). Insets in (D) are zoomed-in plots of hysteresis curves. (**E**) Schematic depicting the spin-orientation-dependent charge hopping between $WSe_2$ and $CrI_3$, which leads to the excitation helicity-dependent PL intensity in Fig. 2A. See text for detailed description. (**F**) PL spectral linewidth (blue) and intensity (purple) vs applied magnetic field (sweeping from positive to negative) for the LL condition. Broad width is always associated with weak PL intensity.

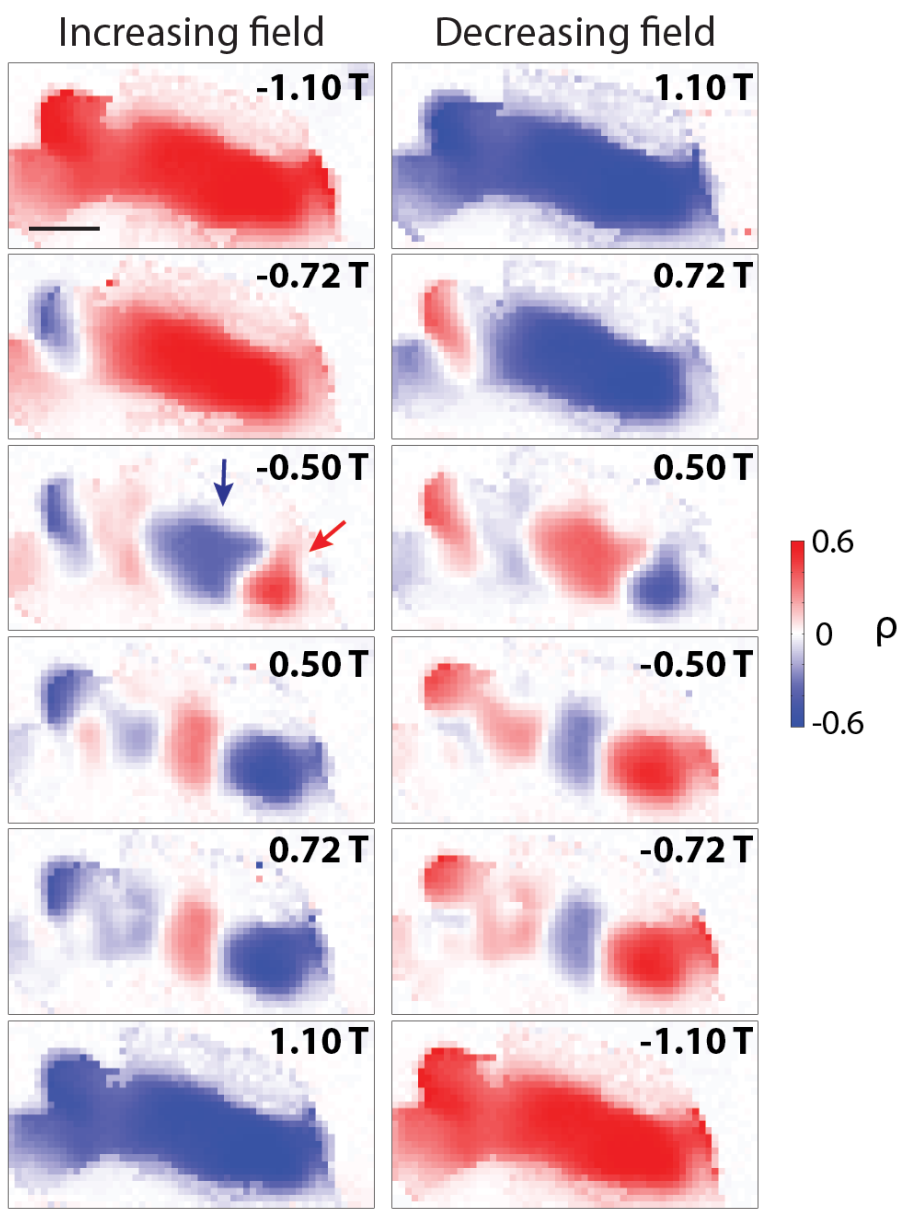

**Fig. 3**. **Polarization-resolved micro-photoluminescence imaging of domain structures.** Each panel is spatial map of ρ (see text for definition) at the indicated applied magnetic field. Left and right columns are arranged in a time-reversal manner corresponding to increasing (left) and decreasing (right) applied magnetic field, respectively. The blue arrow indicates a domain in which the sign of ρ flips three times by sweeping the field, while the red arrow points to a domain that flips the sign of ρ only once. Scale bar, 3 μm.

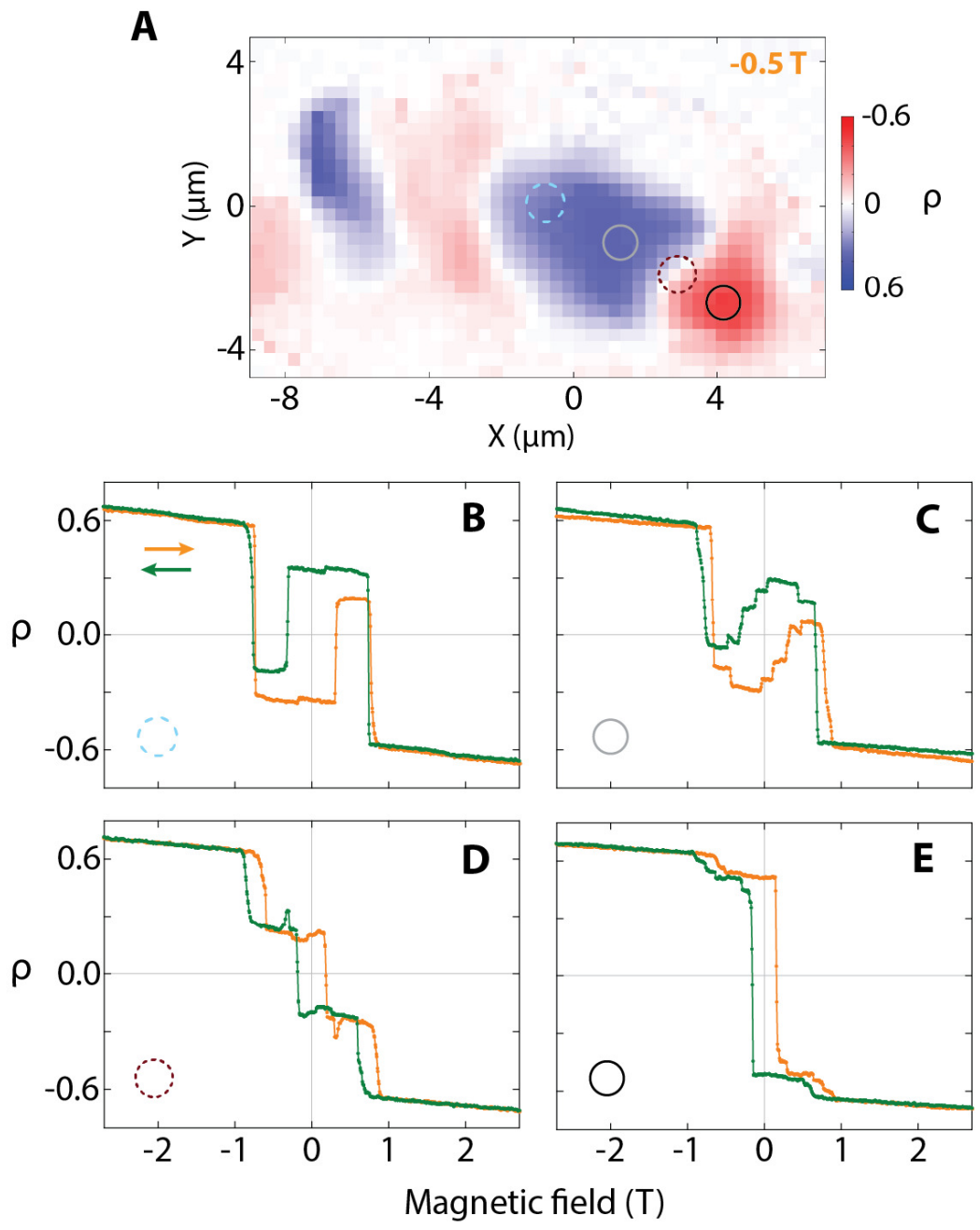

**Fig. 4**. **Position-sensitive ferromagnetic domain dynamics.** (**A**) Spatial map of ρ from Fig. 3 (-0.5 T, sweep down) with blue, gray, brown, and black colored circles indicating the spots of selected magnetic field sweeps of ρ in (**B**), (**C**), (**D**), and (**E**).

# Supplementary Materials for

## Van der Waals Engineering of Ferromagnetic Semiconductor Heterostructures for Spin and Valleytronics

Ding Zhong[†], Kyle L. Seyler[†], Xiayu Linpeng[†], Ran Cheng, Nikhil Sivadas, Bevin Huang, Emma Schmidgall, Takashi Taniguchi, Kenji Watanabe, Michael A. McGuire, Wang Yao, Di Xiao, Kai-Mei C. Fu, Xiaodong Xu[*]

[†]These authors contributed equally to this work.
[*]Correspondence to: xuxd@uw.edu

**Content:**

## S1. Electronic structure of $CrI_3/WSe_2$ bilayer

To determine the band alignment of the $CrI_3$-$WSe_2$ heterostructure, we performed first-principles electronic structure calculations for a bilayer made of a monolayer of $CrI_3$ and a monolayer of $WSe_2$. We used the generalized gradient approximation in the parametrization of Perdew, Burke, and Enzerhof (*44*, *45*), as implemented in the QUANTUM ESPRESSO simulation package (*46*). A vacuum slab of 15 Å was used. An energy cutoff of 130 Ry and a 12×12×1 Monkhorst-Pack special *k*-point mesh for the Brillouin zone integration was found to be sufficient to obtain convergence. To obtain a commensurate heterostructure, 3% strained 2x2 in-plane superlattice of $WSe_2$ ($a_0$ = 3.32 Å) was lattice- matched to $CrI_3$ ($a_0$ = 6.84 Å). This leads to a reduction of the $WSe_2$ band gap compared to the unstrained monolayers, but it does not affect the type of the band alignment. Structural optimization for the bilayer was performed by fixing the in-plane lattice constants to that of the theoretical bulk $CrI_3$ lattice constants. Semiempirical Grimme's DFT-D2 method was used to describe van der Waals interactions (*47*, *48*). The relaxation of the ions was done with the electronic degrees of freedom accurate up to $10^{-6}$ eV. The results presented here are for a stacking configuration where one of the magnetic $Cr^{3+}$ ions is directly below a W atom (Fig. S1A and B).

The orbital projected electronic structure of the bilayer is shown in Fig. S1C. The conduction band minima are mainly made of unoccupied Cr *d*-orbitals with their spin orientation the same as the occupied Cr *d*-orbitals. The valence band maxima (VBM) are mainly made of W $d_{xy}$ and $d_{x^2-y^2}$ orbitals. There is also some hybridization with the I and Se *p*-orbitals in the conduction and valence bands, respectively (not shown). Thus, the valence and conduction band edges are in the two different monolayers, resulting in a type-II band alignment. Furthermore, we have calculated the work functions for monolayer $WSe_2$ and monolayer and bilayer $CrI_3$. Our calculations predict type-II band alignment in both the cases. As the reduction in the band gap from bilayer to bulk $CrI_3$ is only 0.05 eV, we expect the type-II band alignment to persist even for bulk $CrI_3$.

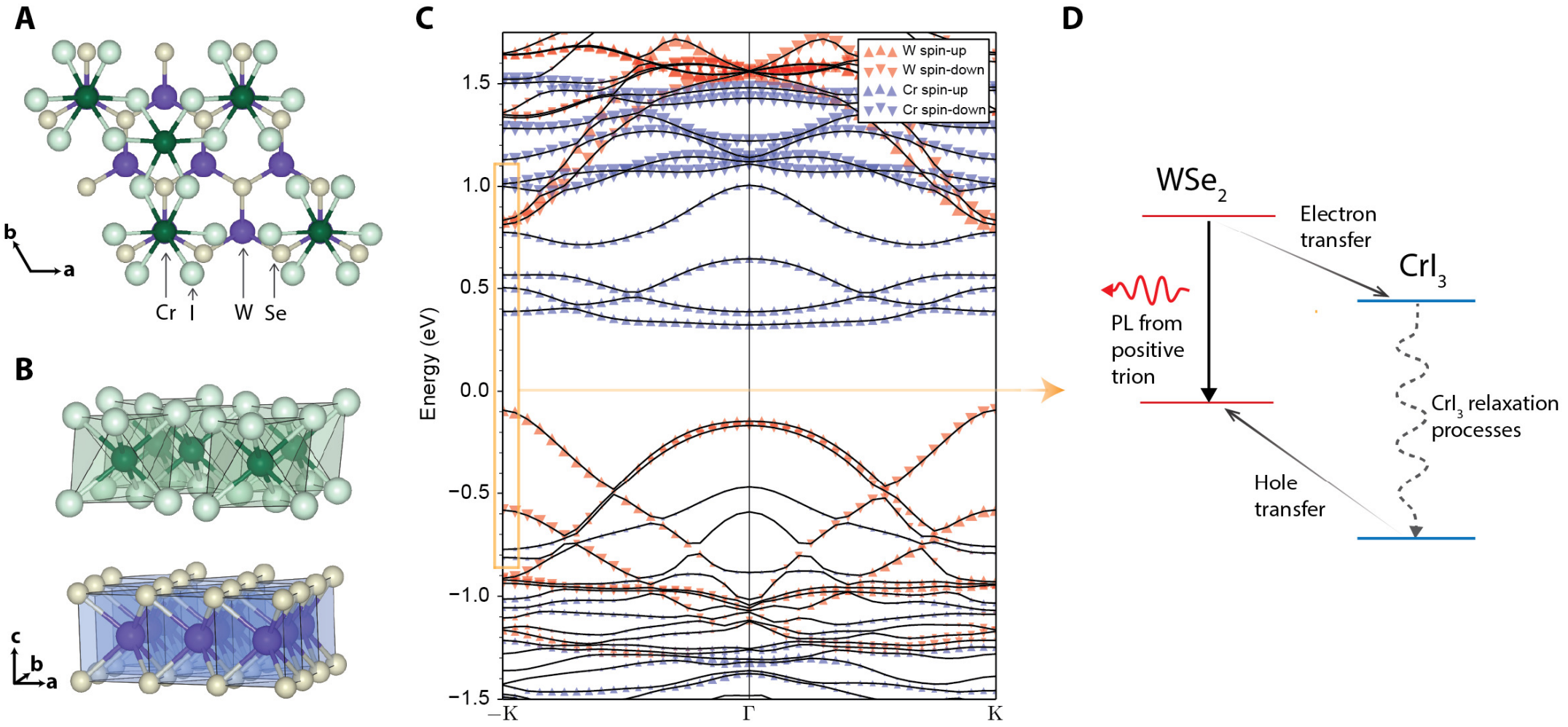

**Fig. S1. The atomic structure and the electronic band structure of the $CrI_3$-$WSe_2$ bilayer**. (**A**) Top and (**B**) side view of the $CrI_3$-$WSe_2$ bilayer. Dark green, light green, purple and tan spheres represent Cr, I, W and Se atoms, respectively. (**C**) Electronic structure of the $CrI_3$-$WSe_2$ bilayer. The contribution from the W (Cr)-orbitals is shown with red (blue) triangles with the triangles pointing up (down) for up (down)-spins. (**D**) Schematic of type-II band alignment. Charge transfer to $CrI_3$ dopes $WSe_2$ *p*-type and PL is thus primarily observed from positively charged trions at the $WSe_2$ ±K valleys.

## S2. Linear polarization

In the main text, we attribute the main peak around 1.68 eV to the trion peak. Multiple pieces of evidence support this assignment: (1) the type-II band alignment between $CrI_3$ and $WSe_2$ (supported by calculations in S1 and magnetic field dependence of PL in Fig. 2) will strongly dope $WSe_2$ with holes; (2) the shoulder peak on the higher energy side of the main peak implies a weak exciton emission. Another compelling piece of evidence is observed in the absence of linear polarization in the main peak. We excited the sample with a linearly polarized 1.96 eV laser while selectively detecting co- ("VV") and cross-linear ("VH") PL (Fig. S2). The beam location is chosen where the higher energy shoulder is prominent. The main peak at 1.681 eV has negligible linear polarization, while the shoulder peak on the high-energy side (1.715 eV) shows clear co-linear polarization. In bare $WSe_2$, while the neutral exciton displays linear polarization due to valley quantum coherence, it is notably absent in the trion emission (*49*). Thus, this strongly supports the assignment of the main peak as trion emission. We also note that the energy separation between the shoulder and main peak is 34 meV, which agrees well with the trion binding energy.

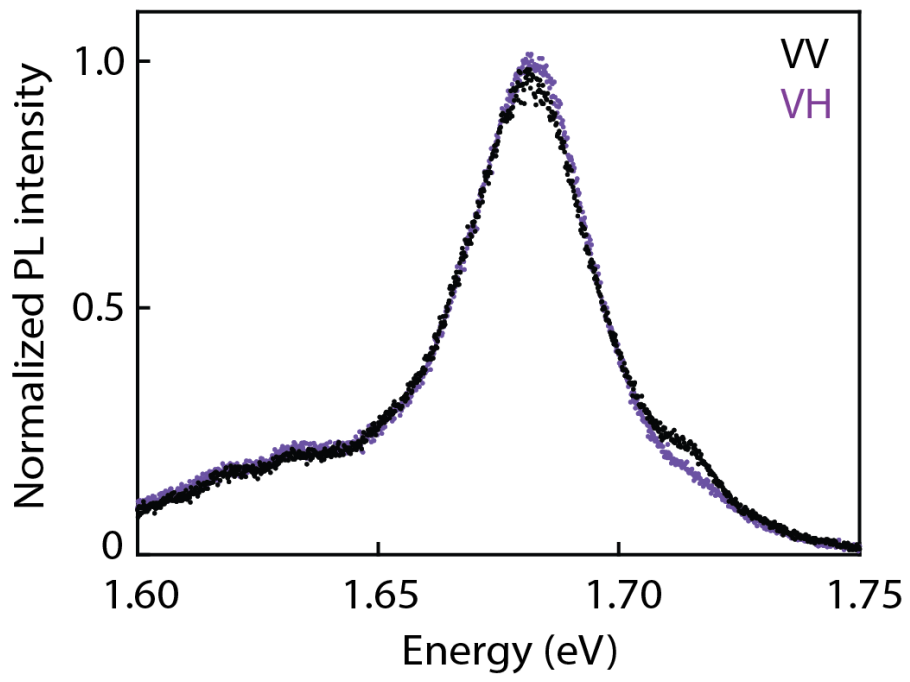

**Figure S2. Linearly polarized excitation and detection.** A comparison of PL intensity between VV (black) and VH (purple) polarization, where the first (second) letter represents excitation (detection) polarization. "H" ("V") stands for horizontal (vertical).

## S3. Peak parameter extraction

We employ two approaches, weighted average and Gaussian fitting, for extracting the parameters of the PL spectrum and compare their results. For the weighted average, the energy of each data point within the full-width at half maximum (FWHM, shaded region in Fig. S3A) were weighted by their PL spectral density, i.e. $\int Ef(E)\, dE \,/ \int f(E) dE$ where $f(E)$ is the spectral density and $E$ is the photon energy. The intensity is defined by $S$/FWHM, where $S$ is the summation of the spectral density within the FWHM.

For the Gaussian fitting, we use multiple Gaussian functions. As shown by the red curve in Fig. S3A, there are weak spectral features on both sides of the main peak. To achieve best fits, we fit the main peak and the high-energy peak with Gaussians. The fitting range is constrained to 1.665 eV on the low energy end to reduce the influence of the low-energy shoulder, which likely does not have a clear peak shape. The results of this fitting scheme are shown in Fig. S3B with the black curve. The dashed curves give the individual Gaussian peaks and blue shows the residual. We see the main peak is captured well by this fitting scheme. The wavy residual between 1.67 and 1.69 eV is due to slight etaloning from the CCD and does not significantly affect the extracted parameters.

Figures S3C and D compare the extracted valley splitting by the two approaches. We see that the magnitude of the splitting as a function of applied magnetic field shows good agreement between both parameter extraction methods, while the noise is smaller in the Gaussian fit method compared with weighted average method. On the other hand, ρ is less sensitive to the fitting approaches, as shown in Figs. S3E and S3F. For the magnetic field sweep data (Fig. 2 and 4) and Fig. 3E, we adopted the Gaussian fitting method. However, for the spatial maps in Fig. 3, we used the weighted average method to extract ρ. This is because the sample contains both heterostructure and bare monolayer $WSe_2$ regions, which have different peak shapes. The weighted average approach thus allows a more straightforward comparison over the entire sample region. Furthermore, we note that ρ captures the sign of the $CrI_3$ magnetization and the valley splitting (Fig. 2C and 2D), and ρ is less noisy, so it allows for higher quality spatially resolved magnetization data than spatially dependent valley splitting (Fig. S12).

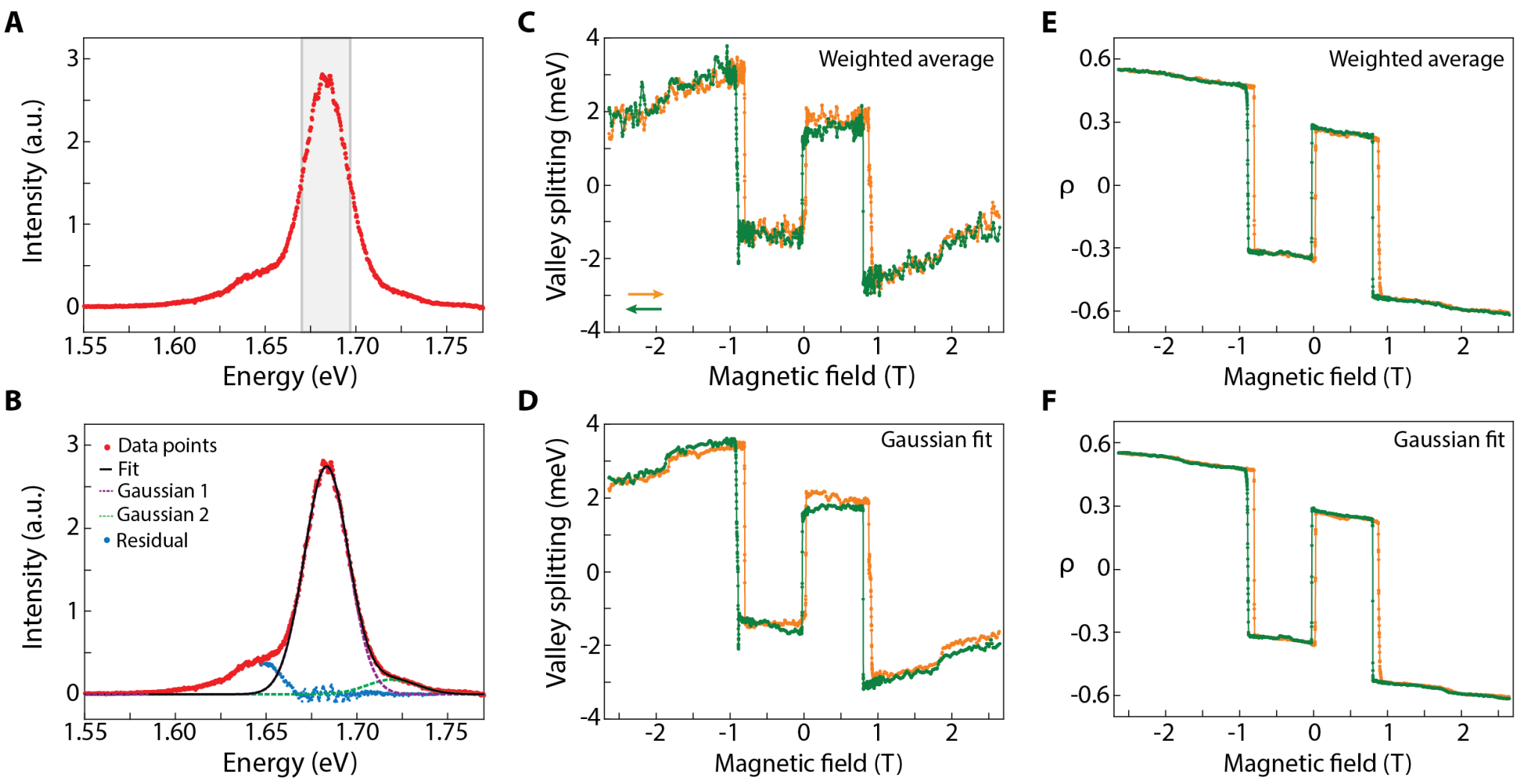

**Figure S3. Peak parameter extraction.** (**A**) PL spectrum of $WSe_2/CrI_3$. The shaded region indicates the points within the FWHM, which are used to calculate the weighted average. (**B**) PL spectrum on $WSe_2/CrI_3$ (red), with Gaussian fits to the main (purple) and high energy (green) peaks. The solid black line represents the total fitted curve and the blue dots give the residual. (**C**) Valley splitting and (**E**) ρ obtained by the weighted average method. (**D**) Valley splitting and (**F**) ρ obtained by Gaussian fitting.

## S4. Comparison of valley splitting between bare $WSe_2$ and $WSe_2/CrI_3$

To estimate the effective magnetic field at the $WSe_2/CrI_3$ interface, we compare the valley splitting of bare $WSe_2$ to the heterostructure. In Fig. S4A, we plot the valley splitting as a function of applied magnetic field for the exciton and trion in $WSe_2$ monolayer. Both the exciton and trion show a similar splitting rate, ~260 μeV/T, which is consistent with previous literature reports on monolayer $WSe_2$ (*31, 32, 37, 38*). The energy splitting between +K and -K is thus approximately -1.5 meV at 5.7 T, as shown in the spectra in Fig. S4B. In comparison, a valley splitting as large as -3.5 meV is observed in $WSe_2/CrI_3$ heterostructure at *zero applied magnetic field* (Fig. S4C).

Using the bare $WSe_2$ valley splitting rate for calibration, we estimate a 13 T effective magnetic field at the $WSe_2/CrI_3$ interface.

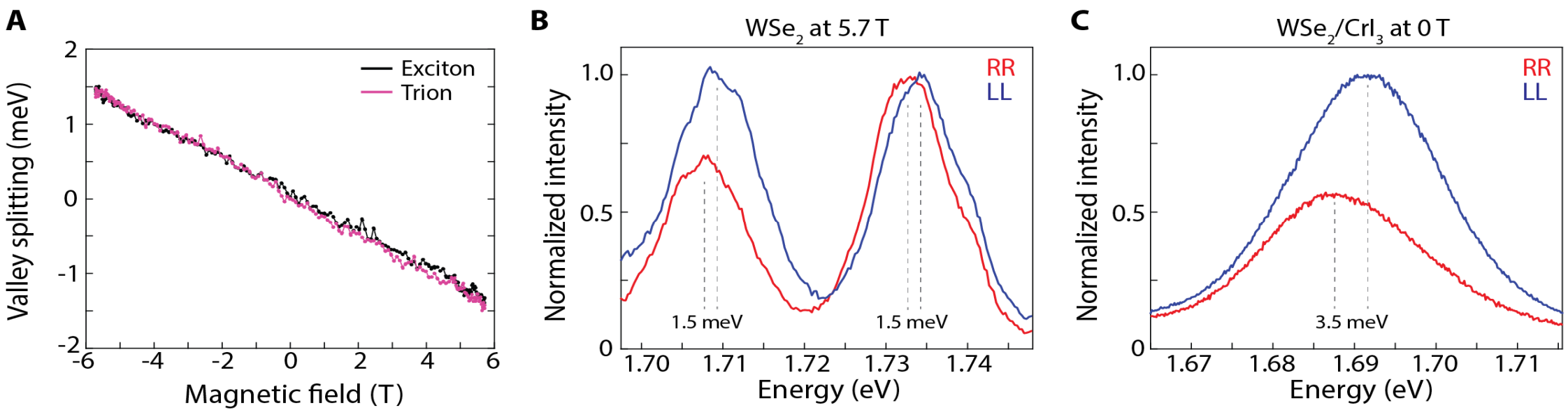

**Figure S4. Valley splitting in bare $WSe_2$ and $WSe_2/CrI_3$.** (**A**) Valley splitting of the exciton (black) and trion (magenta) as a function of applied magnetic field in bare $WSe_2$ monolayer. (**B**) PL spectra for RR (red) and LL (blue) polarization configurations in bare $WSe_2$ at 5.7 T. The valley splitting is approximately -1.5 meV for both the exciton and trion. (**C**) Same as (A) but for the $WSe_2/CrI_3$ heterostructure at 0 T. The valley splitting is measured to be 3.5 meV. The effective magnetic field of the $CrI_3/WSe_2$ interface is estimated to be ~13 T, based on the splitting determined in (A).

## S5. Power dependence of valley splitting

The valley-dependent electron transfer from $WSe_2$ to $CrI_3$ creates a population imbalance between the +K and –K valleys, which may affect their relative energies. In Figure S5, we display the valley splitting under different excitation powers. As the power increases from 1 to 40 μW, the degree of valley splitting remains nearly constant. The independence of the valley splitting on the excitation power implies that carrier density effects are likely not a dominant contributor to the valley splitting. Instead, the exchange interactions between $CrI_3$ and $WSe_2$ are primarily responsible.

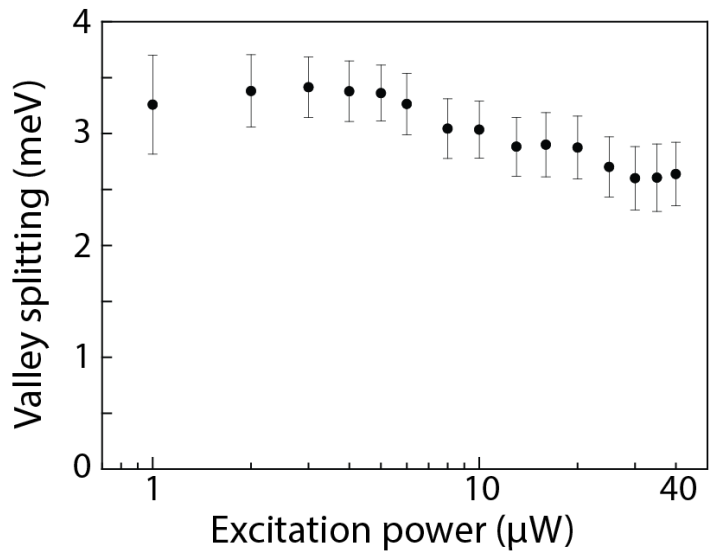

**Figure S5. Power dependence of valley splitting**. Valley splitting as a function of excitation power (log scale), performed at zero applied magnetic field, as in Fig. 1E.

## S6. Voigt geometry

While bulk $CrI_3$ is known to possess an out-of-plane easy axis (*36*), the easy axis in very thin samples has not been established experimentally. In Fig. S6A and S6B, we show the valley splitting and ρ as a function of applied magnetic field in Voigt geometry (field is parallel to the sample plane). A sharp transition that flips the sign of both the valley splitting and ρ occurs near 2.5 T (-2.5 T) when sweeping the field up (down). In addition, we observe a smaller weaker jump occurs near ±4 T, followed by a reduction in the splitting and ρ towards zero at high magnetic fields. The zero crossing in the first jump reflects the magnetization flipping from out-of-plane near zero field towards the opposite direction. At high fields, the valley splitting and ρ approach zero, indicating that the $CrI_3$ magnetization is fully in-plane, aligned with the applied magnetic field. Regardless of the field sweeping direction, the non-vanishing value of the valley splitting and ρ at zero field signifies the existence of out-of-plane magnetization, and the vanishing value of the splitting and ρ at high field indicates an in-plane magnetization, which proves an out-of-plane easy axis.

In addition, we present spatial maps of ρ at selected magnetic fields in Voigt geometry. As shown in Fig. S6C, at 5.7 T, ρ is near zero and no domain structure is visible, consistent with in-plane $CrI_3$ magnetization. At -3 T, we see the presence of domains (Fig. S6D), which increase in strength at zero applied field (Fig. S6E), demonstrating the rotation of the spins towards their preferred out-of-plane direction.

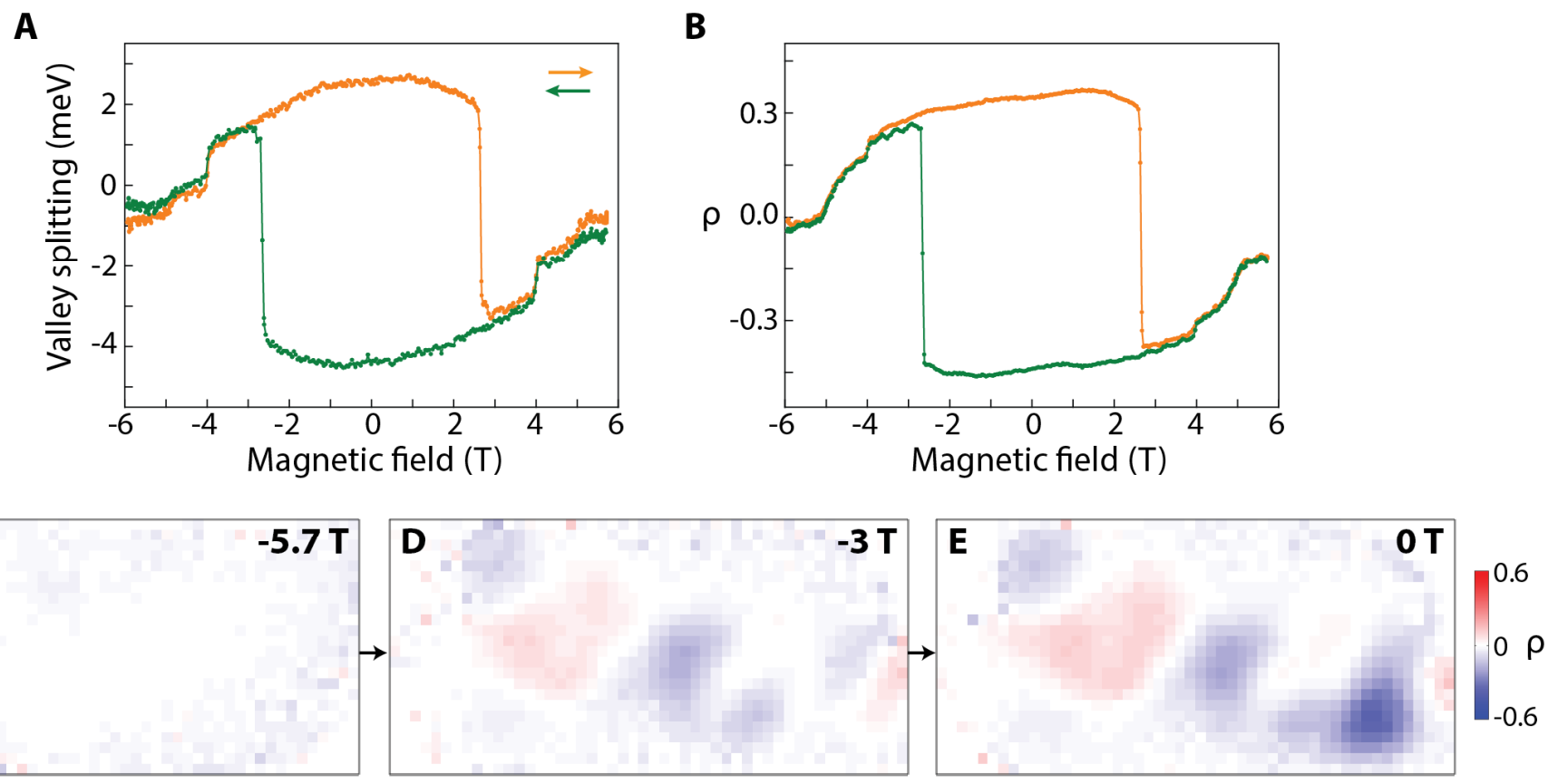

**Figure S6. Photoluminescence measurements in Voigt geometry.** (**A**) Valley splitting and (**B**) ρ versus applied magnetic field, obtained by polarized PL in Voigt geometry. The arrows indicate the sweeping direction. Spatial maps of ρ in Voigt geometry at -5.7 T (**C**), -3 T (**D**), and 0 T (**E**).

## S7. Helicity-independent differential reflection at the excitation energy

We performed polarized differential reflection, $\Delta R/R = (R_{sample} - R_{substrate})/R_{substrate}$), near the excitation energy (1.96 eV) using a tungsten halogen white light source on device 2 at 5 K. As shown in Fig. S7, the $\Delta R/R$ signal, and correspondingly the optical absorption, is independent of the helicity at these energies. This indicates that initial populations created by R or L excitation at 1.96 eV in +K or –K, respectively, are equal and independent of the valley splitting. Thus, the difference in PL intensity under R or L excitation is mainly dominated by valley-dependent non-radiative relaxation channels, as described in the main text.

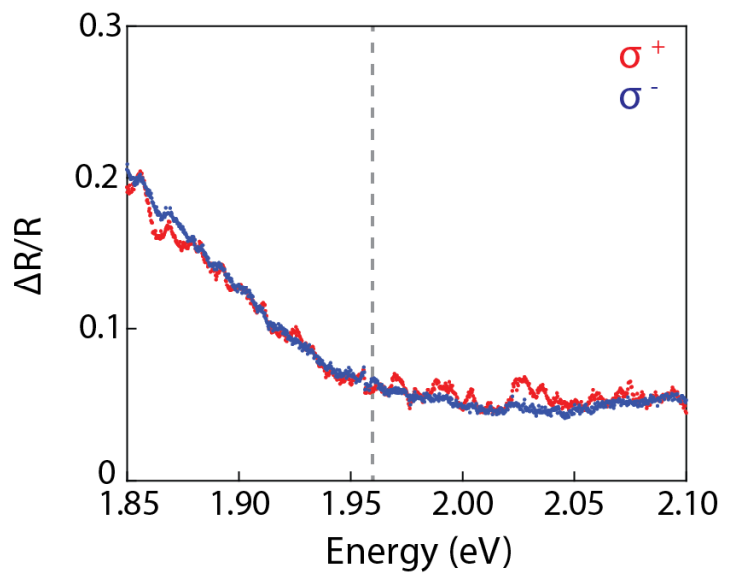

**Figure S7. Differential reflection spectrum of $WSe_2/CrI_3$.** Differential reflection spectrum on $WSe_2/CrI_3$ for σ+ (R, red) and σ- (L, blue) circular excitation. The vertical dashed line indicates the excitation energy of the HeNe laser used in the PL measurements.

## S8. Valley polarization and intensity modulation parameter

The valley polarization is determined by the relative emission intensity between +K and -K after exciting both valleys, i.e. $\frac{I_{RR}+I_{LR}-(I_{RL}+I_{LL})}{I_{RR}+I_{LR}+I_{RL}+I_{LL}}$. In Fig. S8A, the valley polarization is plotted versus applied magnetic field. We see that it displays the same major transitions as the quantity ρ, which we defined in the main text as the normalized difference in RR and LL intensity. Near the transition regions, the valley polarization can switch from ~ 30% to -30% for very small changes in applied magnetic field (mT scale).

It is also useful to study the intensity modulation parameter, $\frac{I_{RR}+I_{RL}-(I_{LR}+I_{LL})}{I_{RR}+I_{RL}+I_{LL}+I_{LR}}$, which contrasts the total PL intensity after exciting +K or -K (Fig. 2A). In Fig. S8B, this parameter is plotted against the applied magnetic field. As discussed in the main text, the highly imbalanced PL after exciting +K or –K reflects the spin-orientation dependent electron transfer to $CrI_3$ based on its magnetization (Fig. 2E). Recently, the phenomenon of spin polarization by spin-dependent charge transfer from a semiconductor to nearby ferromagnetic layer was discovered (*39*). In their system, a GaMnAs layer separated from an InGaAs quantum well, they find a spin polarization and intensity modulation of -3 to 3% from -60 to 60 mT. In contrast, we find that in the $WSe_2/CrI_3$ system, the effect is dramatically enhanced, giving an order of magnitude larger modulation with an order of magnitude smaller magnetic field range.

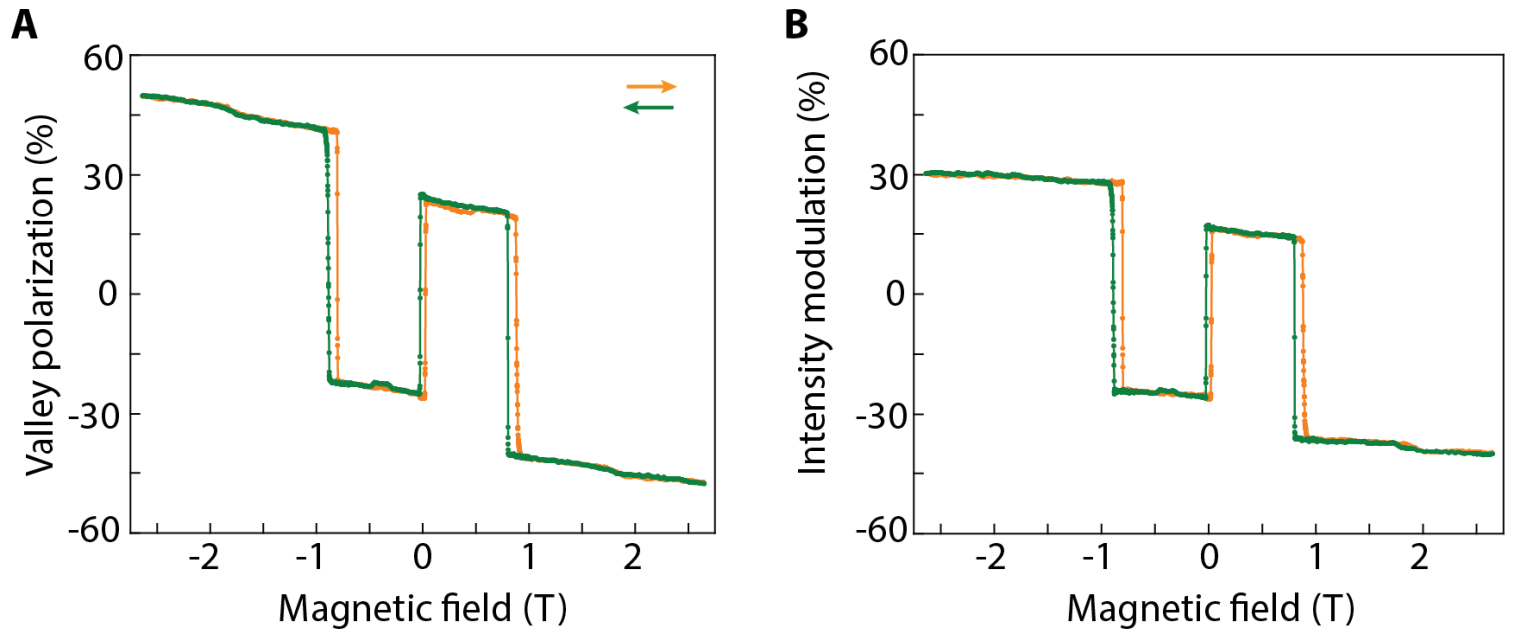

**Figure S8. Valley polarization and intensity modulation parameter. (A)** Valley polarization and **(B)** intensity modulation as function of applied magnetic field. The arrows indicate the sweeping direction of the applied magnetic field.

## S9. Linewidth difference between polarizations

We compare the difference in FWHM between RR and LL spectra as a function of applied magnetic field (Fig. S9). We see that this quantity follows an opposite trend to ρ (Fig. 2D), indicating that the width is large whenever the PL intensity is weak. As noted in the main text, the width difference allows us to estimate the electron hopping rate, which is on the order of 100 fs when $CrI_3$ is fully magnetized. This nonradiative relaxation channel is an order of magnitude faster than the trion radiative lifetime, and therefore has a significant impact on the PL quantum yield. We note that the spikes near ±0.9 T highlight the sensitivity of the spin/valley states in $WSe_2$ to $CrI_3$ magnetization dynamics, and their origin is an intriguing topic for further study.

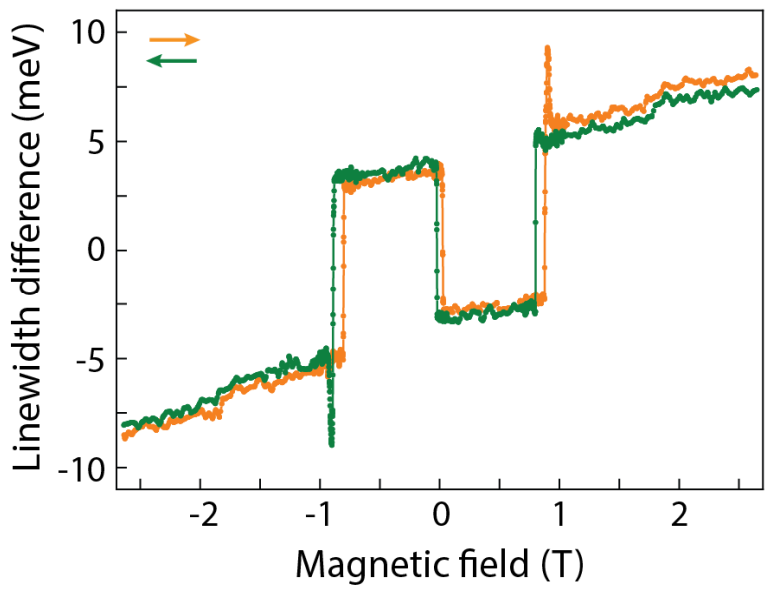

**Figure S9. Linewidth difference between polarizations.** The difference in FWHM as a function of applied magnetic field. Orange and green curves are data for increasing and decreasing applied magnetic field, respectively.

## S10. Rapid switching of heterostructure PL

In Fig. S10, we compare the rate of PL change between plain $WSe_2$ and $WSe_2/CrI_3$. Figures S10A and C are PL spectra maps for LL polarization in plain $WSe_2$ and the HS, respectively. The magnetic field (sweeping up) ranges from -5.7 to 5.7 T in Fig. S10A, whereas it is only plotted around a 40 mT range of the 0.8 T transition in Fig. S10C. This emphasizes the striking contrast in the rate of PL intensity modulation between bare $WSe_2$ and the $WSe_2/CrI_3$ HS. Figures S10B and D illustrate this further with the spectra at the selected field values. The change in spectral position, in addition to intensity, is clear. In a span of over 11 T, there is a 1.5 meV shift in the trion peak position in plain $WSe_2$ and 25% reduction in the PL intensity. For $WSe_2$ interfaced with $CrI_3$, we find a 4 meV shift and 50% reduction in $\leq$ 6 mT.

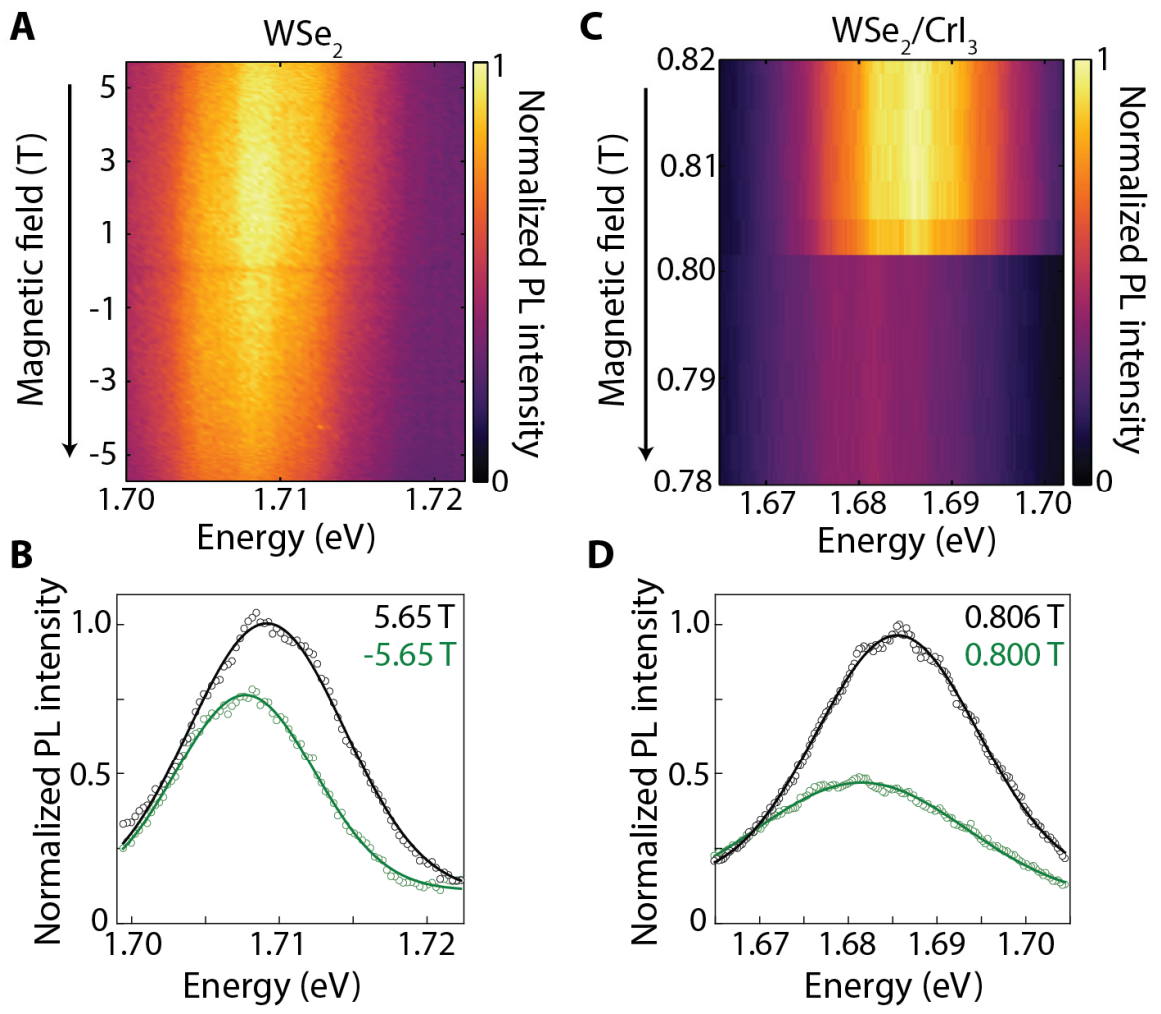

**Figure S10. Rapid switching of PL in $WSe_2/CrI_3$. (A)** Intensity map of LL PL versus applied magnetic field for bare $WSe_2$. **(B)** PL spectra for LL at ±5.65 T. **(C)** Intensity map of LL PL versus applied magnetic field for bare $WSe_2/CrI_3$. **(D)** PL spectra for LL at 0.8 and 0.806 T.

## S11. Magnetic field sweep rate dependence

In Fig. S11, we display the PL intensity (of LL polarization scheme) as a function of decreasing applied magnetic field (from 2.7 to -2.7 T) for different field sweep rates. The fast sweep data was acquired when sweeping the magnetic field at 8.5 mT/s over the entire field range. For the slow sweep data, the sweep rate was lowered to 0.95 mT/s around the critical fields observed in the fast sweep, from 0.96 to 0.72 T, 0.1 to -0.1 T, and -0.72 to -0.96 T. The intensities of the two curves overlap well and the critical fields are within 50 mT of each other (the step size of the fast sweep data), which suggests that the sweep rate does not affect the main results.

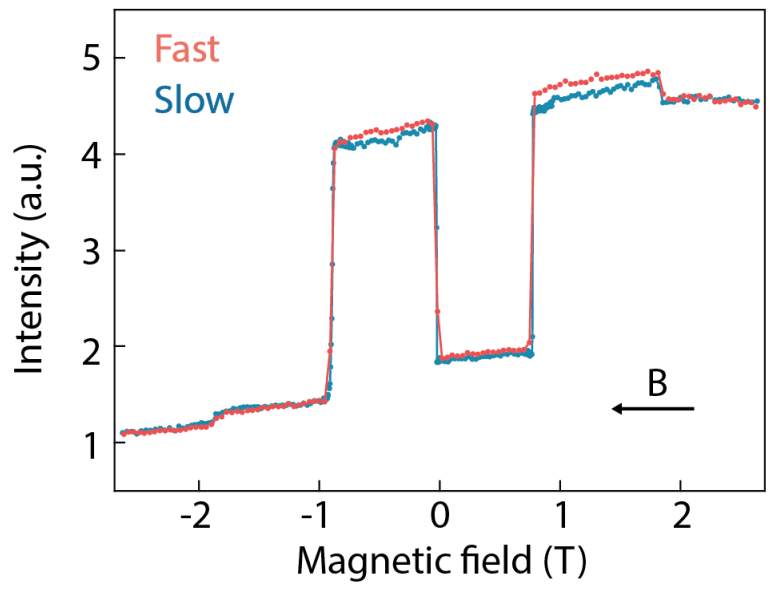

**Figure S11. Sweep rate dependence.** LL PL intensity for the fast sweep is acquired at 8.5 mT/s, while in the slow sweep, the rate is decreased to 0.95 mT/s within ~0.1 T of the critical fields.

## S12. Spatial maps of valley splitting

In Fig. S12, we display the valley splitting spatial maps of device 2. These correspond to the same dataset that was used to extract the spatial maps of ρ in Fig. 3. To avoid fitting issues due to the variation of PL peaks on and off the HS region, we choose to fit only the data where the HS $X^+$ emission has the highest intensity and all other splitting values are set to zero. The agreement of the shape in Fig. S12 to Fig. 3 indicates that this fitting scheme captures the HS area well. Furthermore, the size, shape, and sign of the domain patterns match those in Fig. 3, further justifying the use of ρ as a good parameter to capture the underlying magnetization. In addition, ρ presents a few noteworthy benefits in the data analysis. To extract the valley splitting with lowest uncertainty, one must use a fitting scheme, while for ρ, both fitting and weighted averaging work well (as shown in Fig. S3). Thus, in general, ρ has lower noise than the valley splitting, especially with shorter integration times (1 second per polarization in these spatial maps). Finally, ρ is less sensitive to the peak shapes, allowing extraction of useful results without having to tailor the fits to each pixel.

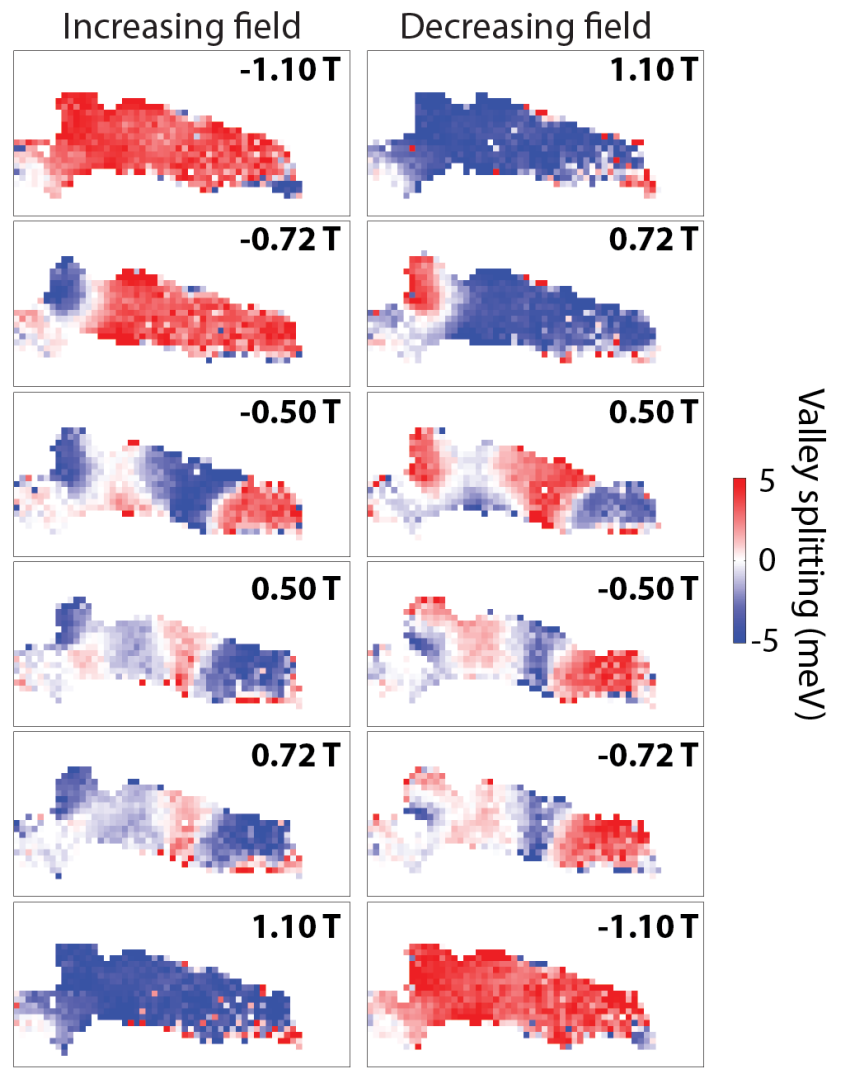

**Figure S12. Spatial maps of valley splitting.** The panels correspond to those in Fig. 3, but the valley splitting on the HS region is plotted instead. Left (right) column arranged in a time-reversal manner corresponding to magnetic field sweeping up (down). See description to the left for fitting details.

## S13. Temperature dependence

Temperature dependence of the ρ spatial maps is shown in Fig. S13. The applied magnetic field is fixed at -0.15 T after sweeping up from -1.1 T. The main feature is the decrease in ρ as temperature increases and its eventual disappearance above ~ 65 K, which suggests that $T_C$ in our ~10 nm $CrI_3$ is comparable to bulk samples (~61 K) (*36*). While the domain structure remains mostly unchanged with temperature, there are subtle changes (e.g. in the shape of the large blue domain), which highlights the utility and sensitivity of the scanning PL technique for probing the $CrI_3$ magnetization.

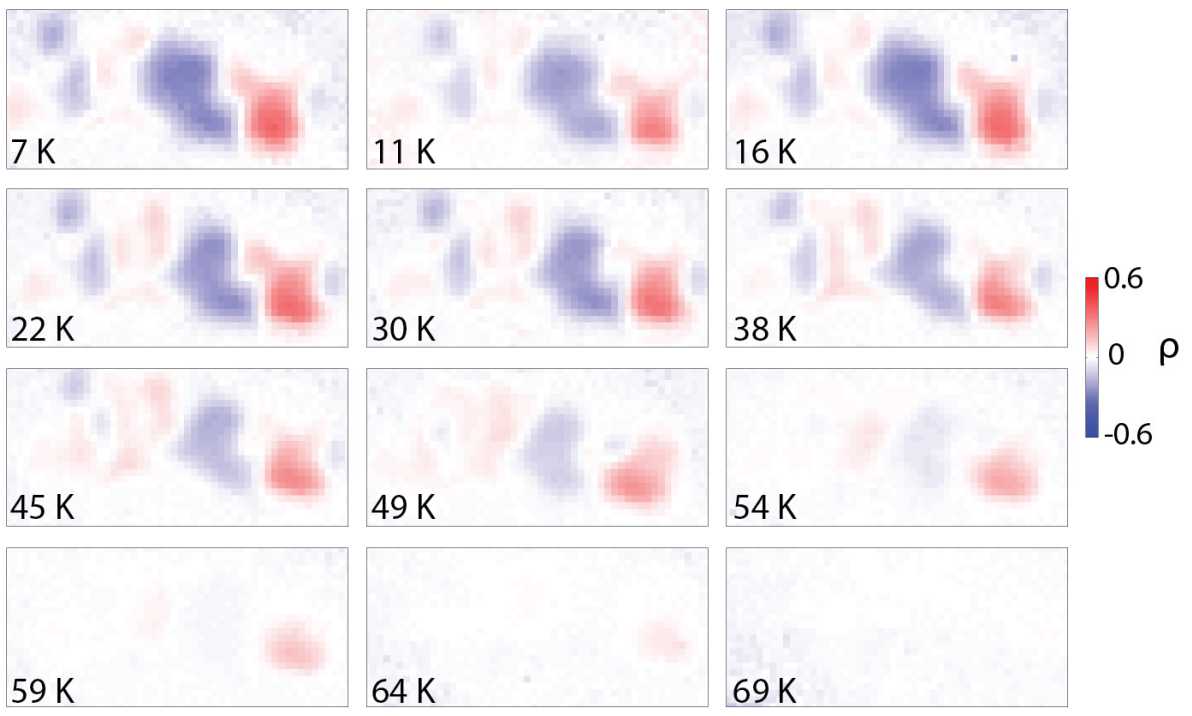

**Figure S13. Temperature dependence of $CrI_3$ magnetization.** Spatial maps of ρ at selected temperatures. The temperatures are indicated on the respective plots.

**<u>S14. Model of strong and weak domain</u>**

The observed PL spectrum reflects the magnetization configuration of the $CrI_3$. In our convention, the absolute value of PL data and the absolute value of $M_z$ (magnetization along the easy-axis) are of opposite sign. In the following, we model the hysteresis loop by focusing on the magnetization direction instead of the PL data, but one should be bear in mind that there is an overall minus sign difference between Fig. S14 and Fig. 2.

Consider a phenomenological model consisting of two monodomains with magnetization vectors denoted by $m_1$ and $m_2$. Let the polar angles of the two domains with respect to the $z$ axis (i.e., the easy-axis) be $\theta_1$ and $\theta_2$. The magnetic switching is realized by individual domain flip, and we ignore DW dynamics induced by the magnetic field as it is not a prominent feature from the data. Nevertheless, the DW energy contributes a term $-D\cos(\theta_1 - \theta_2)$ to the free energy, where D can be determined by system parameters but it suffices to treat D as a phenomenological constant for our purposes. According to the experimental condition, the magnetic field $H$ is parallel to $z$. Assume that $M_1 = m_1V_1 > M_2 = m_2V_2$ where $M_1$ and $M_2$ are the magnetization amplitudes and $V_{1,2}$ are the volumes of the two domains. For either unequal values of magnetization or unequal volumes, the total Zeeman energy of domain 1 is larger than that of domain 2. It is in this sense that we call domain 1 the 'stronger' domain and domain 2 the 'weaker' domain. Our model can be easily generalized to the case of multi-domains, as long as their Zeeman energies are different enough to be divided into strong and weak. The free energy of such a two-domain system is

$$E = (JM_1M_2 - D)\cos(\theta_1 - \theta_2) - H(M_1\cos\theta_1 + M_2\cos\theta_2) + K(M_1\sin^2\theta_1 + M_2\sin^2\theta_2) \quad \text{(S1)}$$

where $K > 0$ is the easy-axis anisotropy, $J > 0$ characterizes the long-range dipolar interaction of two domains (which always favors anti-parallel) and the unit is chosen such that $JM_1M_2$ has an energy dimension. We have neglected all constant terms independent of $\theta_{1,2}$ in Eq. S1 to simplify the following analysis. It is easy to check that for the four configurations listed in Table 1, $\partial E/\partial\theta_1 = \partial E/\partial\theta_2 = 0$, which means that they are all local extrema of the free energy. However, the corresponding second-order derivatives

$$\frac{\partial^2 E}{\partial\theta_1^2} = (D - JM_1M_2)\cos(\theta_1 - \theta_2) + HM_1\cos\theta_1 + 2KM_1\cos 2\theta_1 \tag{S2}$$

$$\frac{\partial^2 E}{\partial\theta_2^2} = (D - JM_1M_2)\cos(\theta_1 - \theta_2) + HM_2\cos\theta_2 + 2KM_2\cos 2\theta_2 \tag{S3}$$

are quite different, indicating that the stability depends on the magnetic field. For example, it is easy to see from Table 1 that in the range $\frac{J}{M_2} - 2K < H < J/M_2 + 2K$, both $(0, 0)$ and $(0, \pi)$ are stable configurations as $\partial^2 E/\partial\theta_1^2 > 0$ and $\partial^2 E/\partial\theta_2^2 > 0$. This range defines a side hysteresis loop of $M_2$ around

$$H_t = \frac{JM_1M_2 - D}{M_2} \tag{S4}$$

as illustrated in Fig S14. Here, $H_t$ marks the point where the Zeeman energy and the dipolar interaction compensate each other in the weak domain.

As long as the inter-domain coupling $J$ is sufficiently large (here $J$ is the dipolar interaction), there are three separate loops based on energy and stability analysis (*50*). The coercivity fields of the center loop and the side loops are, respectively,

$$H_{c1} = 2K\frac{\left(1 + \frac{M_2}{M_1}\right)}{1 - \frac{M_2}{M_1}} \tag{S5}$$

$$H_{c2} = 2K \tag{S6}$$

The actual measured hysteresis loop is quite sensitive to the position of the laser spot. If the spot is placed on a strong domain, the Zeeman energy is always dominant, so it exhibits normal hysteresis loop similar to what is observed in Fig 4E. If the spot is positioned on a weak domain, however, the local magnetic configuration being monitored is determined by the competition between the dipolar $J$ and the Zeeman energy, with a coercivity field set by the anisotropy. If the laser spot covers the boundary between two adjacent domains (Fig. 4D), things become complicated and cannot be captured by a simple model ignoring DW dynamics.

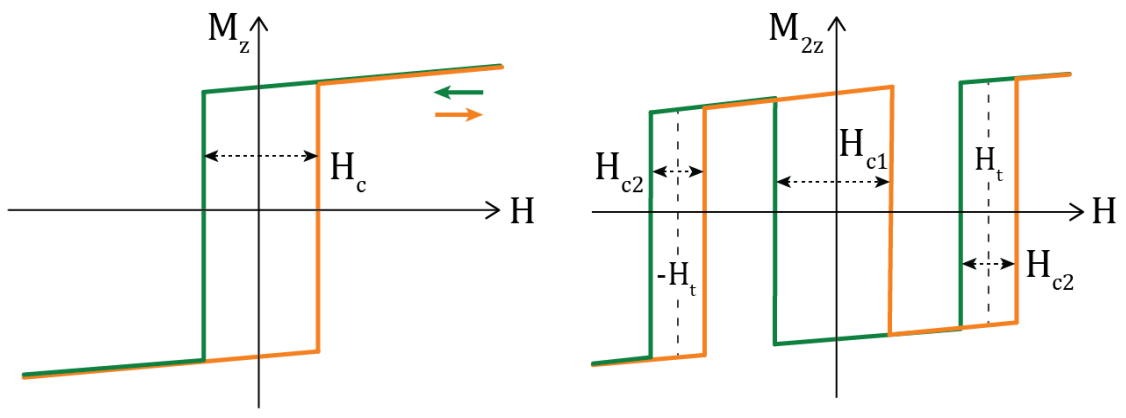

**Fig. S14. Strong and weak domain modelling**. Normal and triple-hysteresis loops. Normal loop corresponds to PL data from strong domains, while triple loops reflects PL data from weak domains.

| $\theta_1$ | $\theta_2$ | $E$ | $\partial^2 E/\partial\theta_1^2$ | $\partial^2 E/\partial\theta_2^2$ |
|---|---|---|---|---|
| 0 | 0 | $(JM_1M_2 - D)$ $- H(M_1 + M_2)$ | -$(JM_1M_2 - D) +$ $HM_1 + 2KM_1$ | $-(JM_1M_2 - D)$ $+ HM_2 + 2KM_2$ |
| 0 | $\pi$ | $-(JM_1M_2 - D)$ $- H(M_1 - M_2)$ | $(JM_1M_2 - D)$ $+ HM_1 + 2KM_1$ | $(JM_1M_2 - D)$ $- HM_2 + 2KM_2$ |
| $\pi$ | 0 | $-(JM_1M_2 - D)$ $+ H(M_1 - M_2)$ | $(JM_1M_2 - D)$ $- HM_1 + 2KM_1$ | $(JM_1M_2 - D)$ $+ HM_2 + 2KM_2$ |
| $\pi$ | $\pi$ | $(JM_1M_2 - D)$ $+ H(M_1 + M_2)$ | $-(JM_1M_2 - D)$ $- HM_1 + 2KM_1$ | $-(JM_1M_2 - D)$ $- HM_2 + 2KM_2$ |

**Table S1.** The free energy and its second-order derivatives with respect to the angles of the two domains. Assume that $JM_1M_2 > D$.